\documentclass[preprint,showpacs,amsmath,amssymb,aps,prd,,nofootinbib]{revtex4}
\usepackage{epsfig}
\begin{document}
\draft
\title{\mbox{}\\[10pt]
Leptogenesis in a seesaw model \\
with Fritzsch type lepton mass matrices}

\author{Y. H. Ahn$^{1,}$\footnote{E-mail:
        yhahn@cskim.yonsei.ac.kr},
        Sin Kyu Kang$^{2,}$\footnote{E-mail:
        skkang@snut.ac.kr},
        C. S. Kim$^{1,}$\footnote{E-mail:
        cskim@yonsei.ac.kr} and
        Jake Lee$^{1,}$\footnote{E-mail:
         jilee@cskim.yonsei.ac.kr}}

\address{$^{1}$  Department of Physics, Yonsei
        University, Seoul 120-749, South Korea\\
        $^{2}$ School of Liberal Arts, Seoul National University of Technology,
        Seoul 139-743, South Korea}

%\thanks{}


\date{\today}

%%%%%%%%%%%%%%%%%%%%%%%%%%%%%%%%%%%%%%%%%%%%%%%%%%%%%%%%%%%%%%%%%%%%%%%%%%%%%%%%%%%%%%%%%%
\begin{abstract}

\noindent
We investigate how the baryon asymmetry of our universe  via leptogenesis can be achieved
within the framework of the seesaw model with Fritzsch type lepton mass matrices proposed
by Fukugita $et.~al$.
We study the cases  with CP-violating phases
in  charged lepton Yukawa matrix, however, with and without Dirac neutrino Yukawa phases.
We consider both flavor independent and flavor dependent leptogenesis, and demonstrate how they lead to
different amounts of lepton asymmetries in detail.
In particular, it is shown that flavor dependent leptogenesis in this model can be worked
out only when the CP phases in Dirac neutrino Yukawa matrix become zero at the GUT scale.
In addition to the CP phases, for successful leptogenesis in  the model
it is required that the degeneracy of the heavy Majorana neutrino mass spectrum should be broken and
we also show that the breakdown of the degeneracy can be radiatively induced.

\end{abstract}
 \maketitle

%%%%%%%%%%%%%%%%%%%%%%%%%%%%%%%%%%%%%%%%%%%%%%%%%%%%%%%%%%%%%%%%%%%%%%%%%%%%%%%%%%%%%%%%%%
\section{Introduction}

Recent precise neutrino experiments appear to show robust evidence
for the neutrino oscillation. The present neutrino experimental
data \cite{atm,SK2002,SNO} exhibit that the atmospheric neutrino
deficit points toward a maximal mixing between the tau and muon
neutrinos. However, the solar neutrino deficit favors a
not-so-maximal mixing between the electron and muon neutrinos. In
addition, although we do not have yet any firm evidence for the
neutrino oscillation arisen from the 1st and 3rd generation flavor
mixing, there is a bound on the mixing element $U_{e3}$ from CHOOZ
reactor experiment, $|U_{e3}|<0.2$ \cite{chooz}. Although
neutrinos have gradually revealed their properties in various
experiments since the historic Super-Kamiokande confirmation of
neutrino oscillations \cite{atm}, properties related to the
leptonic CP violation are completely unknown yet. To understand
the neutrino mixings observed in various oscillation experiments
is one of the most interesting issues in particle physics. The
phenomenon of lepton flavor mixing can be described by a $3\times
3$ unitary matrix $U$, the Maki-Nakagawa-Sakata (MNS) matrix
\cite{MNS}, which contains three mixing angles ($\theta_{12}$,
$\theta_{23}$, $\theta_{13}$) and three CP-violating phases
($\delta$, $\rho$, $\sigma$). Four of these six parameters ($i.e.$,
$\theta_{12}$, $\theta_{23}$, $\theta_{13}$ and $\delta$),
together with two neutrino mass-squared differences ($\Delta
m_{21}^2 \equiv m_2^2 - m_1^2$ and $\Delta m_{32}^2 \equiv m_3^2 -
m_2^2$), can be extracted from the measurements of neutrino
oscillations. At present, a global analysis of current
experimental data yields \cite{Maltoni:2004ei}
  \begin{eqnarray}
   &&0.26\leq \sin^{2}\theta_{12}\leq0.40,~~0.34\leq\sin^{2}\theta_{23}\leq0.67,
   ~~\sin^{2}\theta_{13}\leq0.050\nonumber\\
   &&2.0\leq\Delta m^{2}_{\rm Atm}[10^{-3} {\rm eV}^{2}]\leq2.8,
   ~~7.1\leq\Delta m^{2}_{\rm Sol}[10^{-5}{\rm eV}^{2}]\leq8.3,
  \label{exp bound}
  \end{eqnarray}
%%%---------
at the $3\sigma$ confidence level, but the Dirac CP-violating
phase $\delta$ is entirely unrestricted at present. More accurate
neutrino oscillation experiments are going to determine the size
of $\theta_{13}$, the sign of $\Delta m^2_{32}$ and the magnitude
of $\delta$. The proposed precision experiments for the tritium
beta decay \cite{KATRIN} and the neutrinoless double-beta decay
\cite{0nubeta} will help to probe the absolute mass scale of three
light neutrinos and to constrain the Majorana CP-violating phases
$\rho$ and $\sigma$.

To understand the neutrino mass spectrum and the neutrino mixing
pattern indicated by Eq. (1), Fukugita, Tanimoto and Yanagida
(FTY) have proposed \cite{FTY} an interesting ans$\rm\ddot{a}$tze
to account for current neutrino oscillation data by combining the
Fritzsch texture \cite{Fritzsch} in the seesaw mechanism
\cite{Minkowski} with three degenerate right-handed Majorana
neutrinos. In the FTY ans$\rm\ddot{a}$tze, charged-lepton and
Dirac neutrino Yukawa coupling matrices are also of the Fritzsch
texture, but the heavy Majorana neutrino mass
$\textbf{M}_{R}=M{\bf I}$ with ${\bf I}$ being the $3\times 3$
unit matrix ($i.e.$, $M_i = M$ for $i=1$, 2 and 3) has been assumed.
Then the effective (left-handed) neutrino mass matrix $m_{\rm
eff}$ in the FTY ansatz is no more of the Fritzsch form. Ref.
\cite{FTY} has shown that the FTY ans$\rm\ddot{a}$tze is
compatible very well with current experimental data on solar and
atmospheric neutrino oscillations. And also there have been many
phenomenological analysis \cite{Xing} of FTY model compatible with
current neutrino data.

It is also worthwhile to examine if baryon asymmetry of our
universe (BAU) \cite{cmb} can be viable in the context of FTY
model. In this work, we study how BAU via leptogenesis can be
achieved within the framework of FTY model with possible
CP-violating phases in Dirac neutrino Yukawa matrix and charged
lepton Yukawa matrix. We consider both flavor independent and
dependent leptogenesis, and show how they lead to different
amounts of lepton asymmetries in detail. As will be shown later,
in particular, flavor dependent leptogenesis in the FTY model can
be worked only when the CP phases in Dirac neutrino Yukawa matrix
becomes zero at GUT scale. In addition to the CP phases, for
successful leptogenesis in the FTY model, it is required that the
degeneracy of the heavy Majorana neutrino mass spectrum should be
broken and we show that it can be radiatively induced.

%%%%%%%%%%%%%%%%%%%%%%%%%%%%%%%%%%%%%%%%%%%%%%%%%%%%%%%%%%%%%%%%%%%%%%%%%%%%%%%%%%%%%%%%%%%%%%%%%%%%%
\section{FTY model realized at GUT scale and CP violation}

Let us begin by considering the Standard Model (SM) of the seesaw
mechanism, which is given by
 \begin{eqnarray}
 {\cal L}\supset e_R^{cT} \textbf{Y}_l L\cdot \overline{\varphi}+N^{cT}_R \textbf{Y}_{\nu}L\cdot\varphi
 -\frac{1}{2}N^{cT}_R \textbf{M}_R N^{c}_R+h.c,
 \label{lagrangian}
 \end{eqnarray}
where the family indices have been omitted and
$L_{\alpha}(\alpha=e,\mu,\tau)$ stand for the left-handed lepton
doublets, $(e_R^{c})_{\alpha}$ are the charged lepton singlets,
$N_{R\alpha}$ the right-handed neutrino singlets and  $\varphi$ is
the Higgs doublet fields. In the above lagrangian, $\textbf{Y}_l$ and
$\textbf{Y}_{\nu}$ are the $3\times 3$ charged lepton and neutrino
Dirac Yukawa matrices, respectively. After spontaneous symmetry
breaking, the seesaw mechanism leads to a following effective
light neutrino mass term,
\begin{eqnarray}
  &&m_{\rm eff}=-\textbf{Y}^{T}_{\nu}\textbf{M}^{-1}_{R}\textbf{Y}_{\nu}\upsilon^{2} ~,
  \label{meff}
\end{eqnarray}
where $\upsilon$ is a vacuum expectation value of the Higgs field
$\varphi$  with $\upsilon \simeq 174$ GeV.

Let us assume that the charged-lepton mass matrix $m_l=\upsilon
\textbf{Y}_{l}$ and the Dirac neutrino mass matrix $m_{\rm
D}=\upsilon \textbf{Y}_{\nu}$ are both symmetric and of the
Fritzsch texture, at the high energy scale, where
\begin{eqnarray}
  \textbf{Y}_{l}= {\left(\begin{array}{ccc}
 {\bf 0} &  A_{l}e^{i\varphi_{A}} &  {\bf 0} \\
 A_{l}e^{i\varphi_{A}} &  {\bf 0} &  B_{l}e^{i\varphi_{B}} \\
 {\bf 0} &  B_{l}e^{i\varphi_{B}} &  C_{l}
 \end{array}\right)}~~~\textbf{Y}_{\nu}= {\left(\begin{array}{ccc}
 {\bf 0} &  A_{\nu}e^{i\phi_{A}} &  {\bf 0} \\
 A_{\nu}e^{i\phi_{A}} &  {\bf 0} &  B_{\nu}e^{i\phi_{B}} \\
 {\bf 0} &  B_{\nu}e^{i\phi_{B}} &  C_{\nu}
 \end{array}\right)\;.}
\label{FTY1}
 \end{eqnarray}
Here
$A_{l(\nu)},B_{l(\nu)},C_{l(\nu)},\phi_{A},\phi_{B},\varphi_{A}$
and $\varphi_{B}$ are taken to be all real and positive without loss of
generality and then only the off-diagonal elements of $Y_{l(\nu)}$
are complex. Following the FTY ansatz, we take the right-handed Majorana neutrino mass matrix
to be,
 \begin{eqnarray}
  \textbf{M}_{R}=M{\bf I}.
 \end{eqnarray}
In the basis where the charged lepton Yukawa coupling matrix and
the mass matrix of the right-handed neutrino singlets are
diagonal,
 \begin{eqnarray}
  e_{R}\rightarrow V_{R}e_{R},~~~\nu_{L}\rightarrow V_{L}\nu_{L}\;,
 \end{eqnarray}
and the Yukawa matrices of $\textbf{Y}_{l}$ and $\textbf{Y}_{\nu}$
transform as
 \begin{eqnarray}
  \textbf{Y}_{l}\rightarrow V^{\dag}_{R}\textbf{Y}_{l}V_{L},~~~~\textbf{Y}_{\nu}\rightarrow
  \textbf{Y}_{\nu}V_{L}
 \end{eqnarray}
where $V_{R(L)}$ are the unitary matrices to diagonalize
$\textbf{Y}_{l}$. Since the charged-lepton Yukawa matrix
$\textbf{Y}_{l}$ is symmetric in the present framework, only one
unitary matrix, $V_{L}=V_{R}\equiv V$, is sufficient to
diagonalize $\textbf{Y}_{l}$. Then, the transformed Yukawa
matrices $Y^{\prime}_{l}$ and $Y_{\nu}^{\prime}$ are given by
 \begin{eqnarray}
  \textbf{Y}'_{l}= V^{\dag}\textbf{Y}_{l}V={\left(\begin{array}{ccc}
 Y_{e} &  0 &  0 \\
 0 &  Y_{\mu} &  0 \\
 0 &  0 &  Y_{\tau}
 \end{array}\right)},~~~~
 \textbf{Y}'_{\nu}= {\left(\begin{array}{ccc}
 0 &  A_{\nu}e^{i\phi_{A}} &  0 \\
 A_{\nu}e^{i\phi_{A}} &  0 &  B_{\nu}e^{i\phi_{B}} \\
 0 &  B_{\nu}e^{i\phi_{B}} &  C_{\nu}
 \end{array}\right)}V\;.
 \label{rebasing}
 \end{eqnarray}
In addition,  $\textbf{Y}_{l}$ can be decomposed as
$\textbf{Y}_{l}=P^{T}\widehat{Y}_{l}P$
 with $P={\rm diag}(e^{i(\varphi_{A}-\varphi_{B})},e^{i\varphi_{B}},1)$ and
 \begin{eqnarray}
 \widehat{Y}_{l}= {\left(\begin{array}{ccc}
 0 &  A_{l} &  0 \\
 A_{l} &  0 &  B_{l} \\
 0 &  B_{l} &  C_{l}
 \end{array}\right)\;.}
 \end{eqnarray}
 Then, the mass matrix $\textbf{Y}_{l}$ can  finally  be diagonalized by the unitary matrix
 $V = PO$ where the elements of the orthogonal matrix $O$ can be presented in terms of two parameters
$x\equiv y_{e}/y_{\mu}$ and $y\equiv y_{\mu}/y_{\tau}$ as follows,
 \begin{eqnarray}
  O_{11} &=& +\left [ \frac{1-y}{(1+x)(1-xy)(1-y+xy)} \right]^{1/2},~~~~~~O_{12}
           = -i \left [ \frac{x(1+xy)}{(1+x)(1+y)(1-y+xy)}\right ]^{1/2},\nonumber \\
  O_{13} &=& +\left [ \frac{xy^3(1-x)}{(1-xy)(1+y)(1-y+xy)} \right ]^{1/2},~~~~~~O_{21}
           = + \left [\frac{x(1-y)}{(1+x)(1-xy)} \right ]^{1/2},\nonumber \\
  O_{22} &=& +i\left [ \frac{1+xy}{(1+x)(1+y)} \right ]^{1/2},~~~~~~~~~~~~~~~~~~~~~~~O_{23}
           = + \left [ \frac{y(1-x)}{(1-xy)(1+y)} \right ]^{1/2},\nonumber \\
  O_{31} &=& -\left [ \frac{xy(1-x)(1+xy)}{(1+x)(1-xy)(1-y+xy)}\right ]^{1/2},~~~~~~O_{32}
           =-i \left [ \frac{y(1-x)(1-y)}{(1+x)(1+y)(1-y+xy)}\right ]^{1/2},\nonumber \\
  O_{33} &=& +\left [ \frac{(1-y)(1+xy)}{(1-xy)(1+y)(1-y+xy)}\right ]^{1/2}.
\end{eqnarray}
The Dirac neutrino Yukawa matrix can also be written
in the basis we consider as,
\begin{eqnarray}
 \textbf{Y}'_{\nu}&=& B_{\nu}{\left(\begin{array}{ccc}
 0 &  \omega e^{i\phi_{A}} &  0 \\
 \omega e^{i\phi_{A}} &  0 &  e^{i\phi_{B}} \\
 0 &  e^{i\phi_{B}} &  \kappa
 \end{array}\right)}{\left(\begin{array}{ccc}
 e^{i(\varphi_{A}-\varphi_{B})} &  0 &  0 \\
 0 &  e^{i\varphi_{B}} & 0  \\
 0 &  0 &  1
 \end{array}\right)}{\left(\begin{array}{ccc}
 O_{11} &  O_{12} &  O_{13} \\
 O_{21} &  O_{22} &  O_{23} \\
 O_{31} &  O_{32} &  O_{33}
 \end{array}\right)}
 \end{eqnarray}
where the parameters $\omega$ and $\kappa$ are defined by
\begin{eqnarray}
 \omega\equiv \frac{A_{\nu}}{B_{\nu}},~~\kappa\equiv \frac{C_{\nu}}{B_{\nu}}.
 \label{parameter}
\end{eqnarray}
Then, we are led to the effective light neutrino mass matrix  as follows,
\begin{eqnarray}
 && m_{\rm eff} = -\frac{\upsilon^{2}}{M}\textbf{Y}'^{T}_{\nu}\textbf{Y}'_{\nu}\nonumber\\
  &&= \frac{-B_{\nu}^{2}\upsilon^{2}}{M}O^{T}{\left(\begin{array}{ccc}
 e^{2i(\phi_{A}+\varphi_{A}-\varphi_{B})}\omega^{2} &  0  &  e^{i(\phi_{A}+\phi_{B}+\varphi_{A}-\varphi_{B})}\omega \\
 0 & e^{2i\varphi_{B}}( e^{2i\phi_{B}}+e^{2i\phi_{A}}\omega^{2}) &  e^{i(\phi_{B}+\varphi_{B})}\kappa \\
 e^{i(\phi_{A}+\phi_{B}+\varphi_{A}-\varphi_{B})}\omega &  e^{i(\phi_{B}+\varphi_{B})}\kappa &  e^{2i\phi_{B}}+\kappa^{2}
 \end{array}\right)}O.
  \label{meff}
\end{eqnarray}

Concerned with CP violation, we notice from Eq. (\ref{meff}) that
the CP phases $\phi_{A,B}$ coming from  $\textbf{Y}_{\nu}$ as well
as the CP phases $\varphi_{A,B}$ from $\textbf{Y}_{l}$ obviously
take part in low energy CP violation because low energy CP
violation is associated with the form
$\textbf{Y}'^{T}_{\nu}\textbf{Y}'_{\nu}$. On the other hand,
flavor independent leptogenesis is associated with the form given
by
\begin{eqnarray}
  \textbf{Y}'_{\nu}\textbf{Y}'^{\dag}_{\nu}=\textbf{Y}_{\nu}\textbf{Y}^{\dag}_{\nu}=B^{2}_{\nu}\left(\begin{array}{ccc}
  \omega^{2}   &  0  & \omega e^{i(\phi_{A}-\phi_{B})}  \\
  0  &  1+\omega^{2}   & \kappa e^{i\phi_{B}} \\
  \omega e^{-i(\phi_{A}-\phi_{B})} &  \kappa e^{-i\phi_{B}}  & 1+\kappa^{2}  \\
\end{array}
\right). \label{YYd}
\end{eqnarray}
{}From this, we find that only the phases $\phi_{A},~\phi_{B}$ in
$\textbf{Y}_{\nu}$  take part in leptogenesis. However, the
situation is changed when we consider the scenario of {\it
flavored leptogenesis} \cite{Flavor}, where flavor effects become
important. As will be shown later, the magnitude of CP asymmetry
in the scenario of flavored leptogenesis crucially depends on the
following quantity
\begin{eqnarray}
  &&{\rm Im}\{(\textbf{Y}_{\nu}\textbf{Y}^{\dag}_{\nu})_{jk}(\textbf{Y}_{\nu})_{j\alpha}(\textbf{Y}_{\nu})^{\dag}_{\alpha k}\}\nonumber\\
  &&={\rm Im}[(\textbf{Y}_{\nu}\textbf{Y}^{\dag}_{\nu})_{jk}]{\rm Re}[(\textbf{Y}_{\nu})_{j\alpha}(\textbf{Y}_{\nu})^{\dag}_{\alpha k}]
  +{\rm Re}[(\textbf{Y}_{\nu}\textbf{Y}^{\dag}_{\nu})_{jk}]{\rm Im}[(\textbf{Y}_{\nu})_{j\alpha}(\textbf{Y}_{\nu})^{\dag}_{\alpha k}].
  \label{YYdf}
 \end{eqnarray}
This quantity implies that both CP phases in $\textbf{Y}_{\nu}$
and $\textbf{Y}_{l}$ take part in flavored leptogenesis. Contrary
to the case of flavor independent leptogenesis, flavored
leptogenesis can be realized without the CP phases appeared in
$\textbf{Y}'_{\nu}\textbf{Y}'^{\dag}_{\nu}$ as long as the phases
$\varphi_{A,B}$ are non-zero. In addition, we expect that in the
FTY model, there may exist a connection between flavored
leptogenesis with low energy CP violation, contrary to the observation from the generic
seesaw model with three generations \cite{ellis}.

%%%%%%%%%%%%%%%%%%%%%%%%%%%%%%%%%%%%%%%%%%%%%%%%%%%%%%%%%%%%%%%%%%%%%%%%%%%%%%%%%%%%%%%%%%%%%%%%%%%%%%%%%%%%%%%
\section{Confronting with Low-energy neutrino data}

Before discussing how to achieve leptogenesis in FTY model, we
first examine if it is consistent with low energy neutrino data.
To do so, we need renormalization group (RG) evolution
\cite{RGE,Antusch:2005gp,RG1} of neutrino Dirac-Yukawa matrix and
heavy Majorana neutrino masses with the FTY forms from the GUT
scale to the seesaw scale by numerically solving all the relevant
RG equations presented in Ref. \cite{Antusch:2005gp}. For our
purpose, we consider two cases, one is the case with non-vanishing
CP phases in both $\textbf{Y}_{\nu}$ and $\textbf{Y}_{l}$, $\phi_{A,B}\neq0$ and
$\varphi_{A,B}\neq0$, and the other is the case that only the
phases $\varphi_{A,B}$ are non-zero, $i.e.$ $\phi_{A,B}=0$ and
$\varphi_{A,B}\neq0$. Then, we solve the RGE's by varying input
values of the parameter set$\{B_{\nu}, \kappa, \omega,
\varphi_{A},\varphi_{B},\phi_{A}, \phi_{B}, M\}$, and $\{B_{\nu},
\kappa, \omega,\varphi_{A},\varphi_{B}, M\}$ given at the GUT
scale, respectively, and determine the parameter set which is in
consistent with low energy neutrino data. In our numerical
calculation, we use five experimental results for neutrino mixing
parameters and mass squared differences at $3\sigma$
\cite{Maltoni:2004ei} by Eq. (\ref{exp bound}) as inputs.

\begin{figure}[b]
%\vspace*{-5.0cm}
\hspace*{-2cm}
\begin{minipage}[t]{6.0cm}
\epsfig{figure=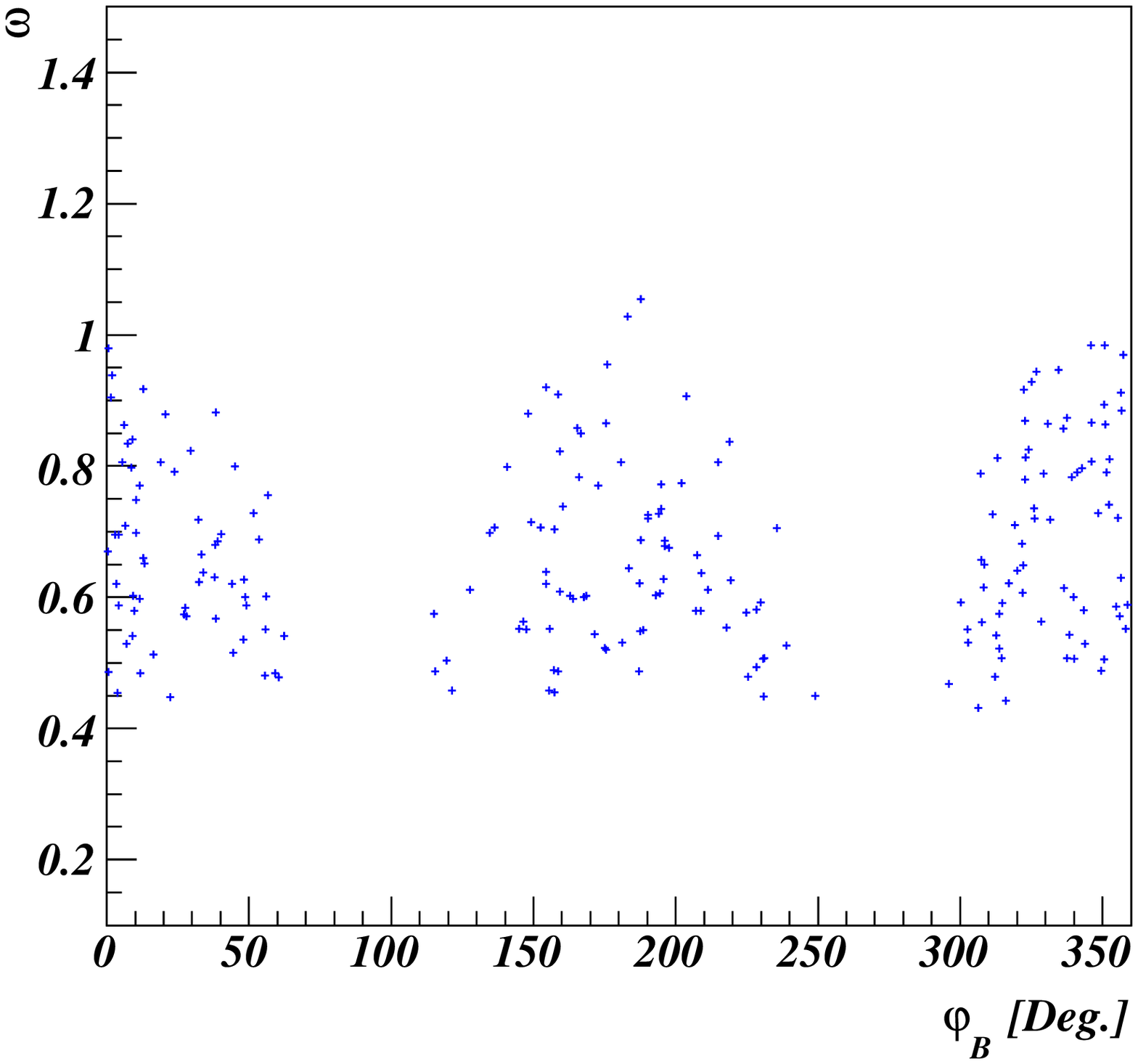,width=6.5cm,angle=0}
\end{minipage}
\hspace*{2.0cm}
\begin{minipage}[t]{6.0cm}
\epsfig{figure=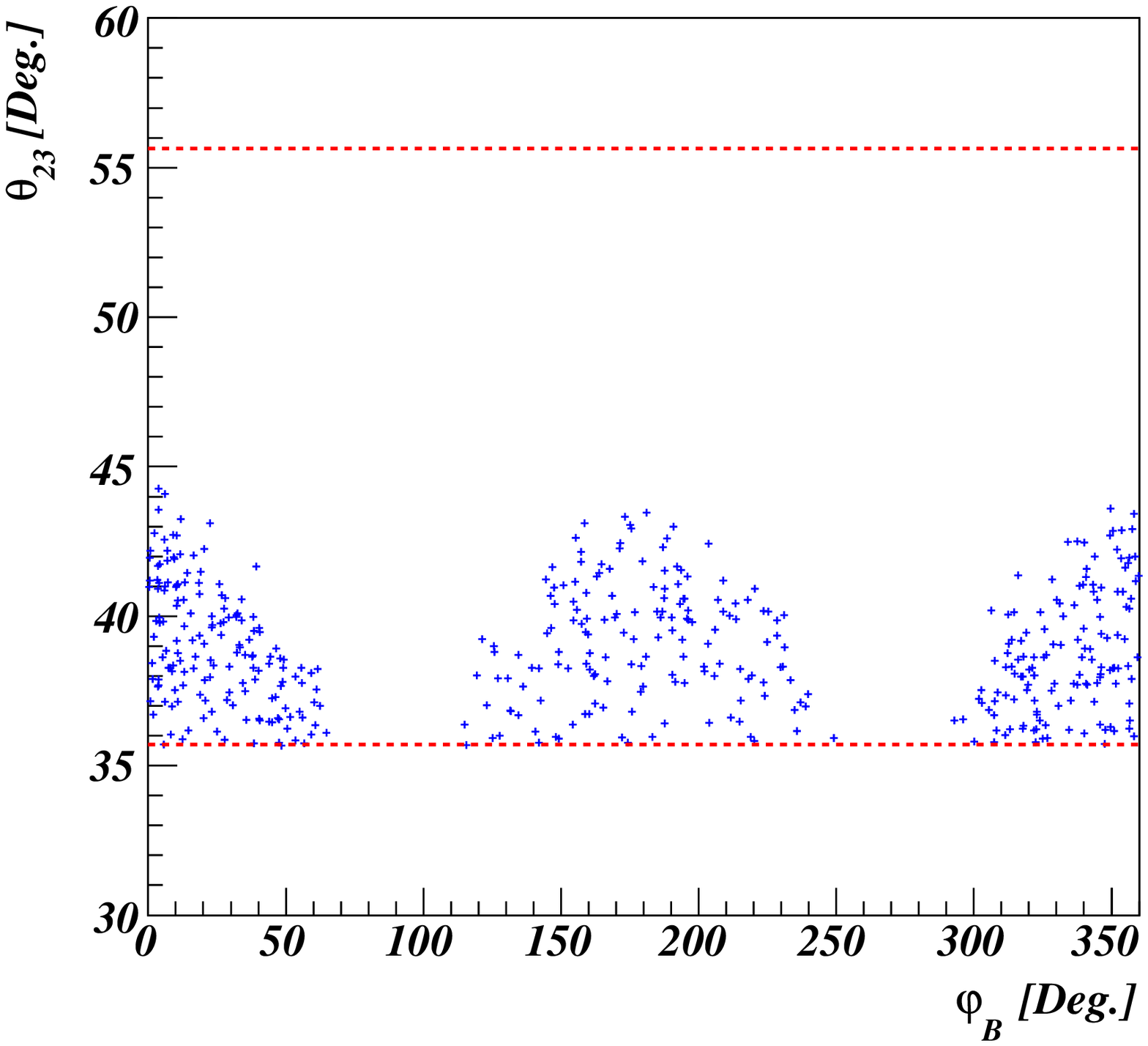,width=6.5cm,angle=0}
\end{minipage}
\vspace*{-1.0cm} \hspace*{-2cm}
\begin{minipage}[t]{6.0cm}
\epsfig{figure=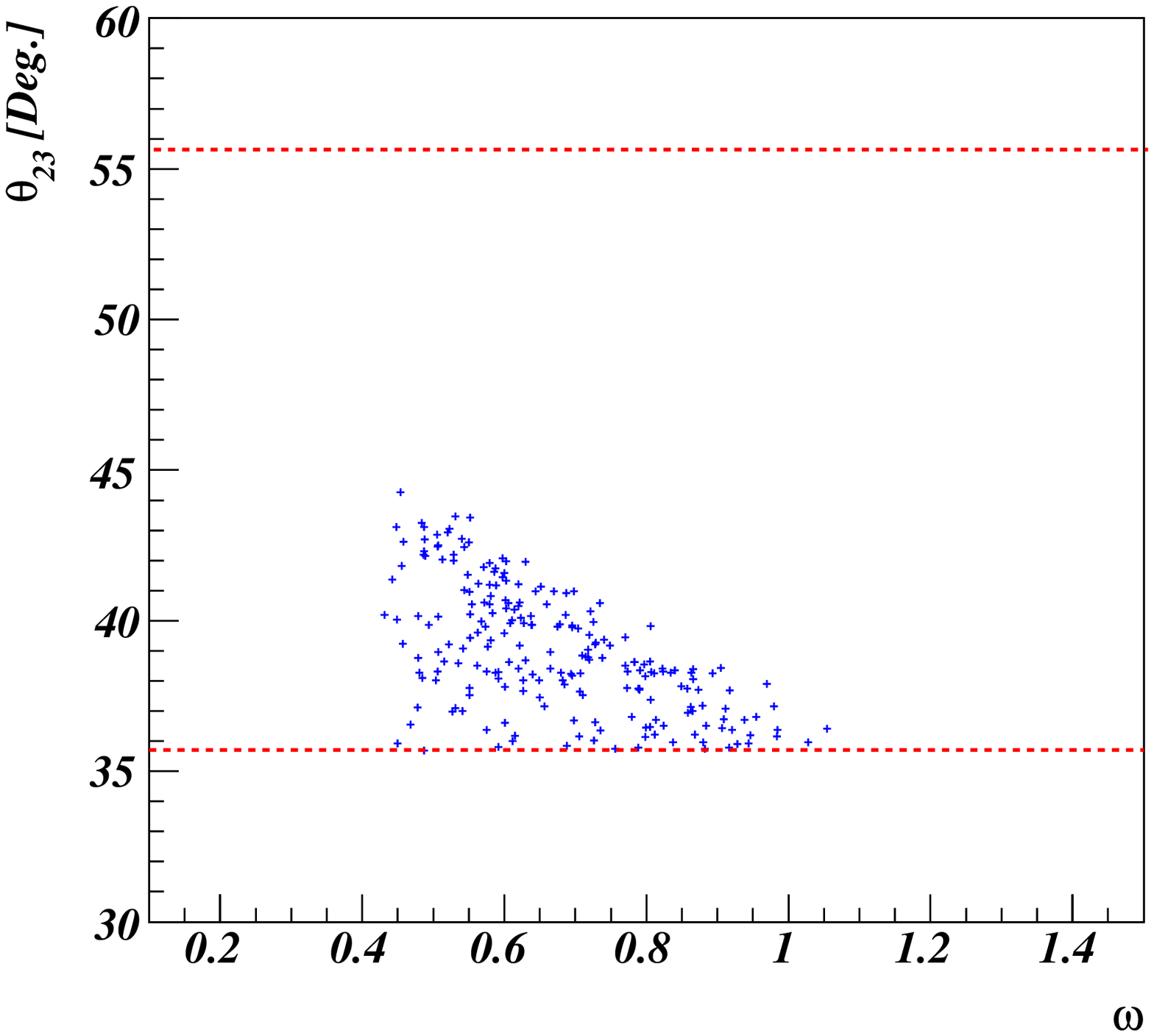,width=6.5cm,angle=0}
\end{minipage}
\hspace*{2.0cm}
\begin{minipage}[t]{6.0cm}
\epsfig{figure=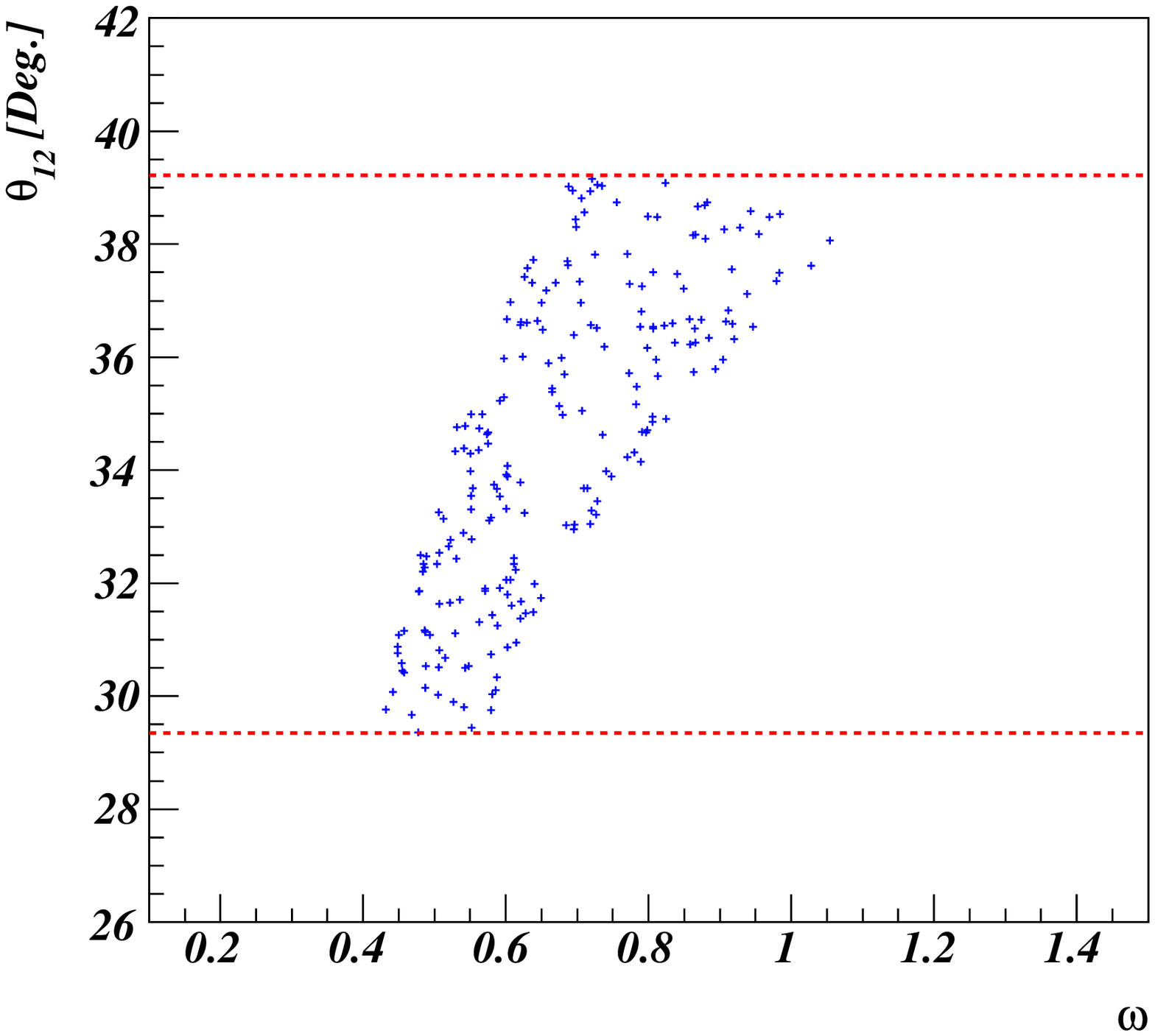,width=6.5cm,angle=0}
\end{minipage}
%\vspace*{-1.0cm}
\caption{\label{Fig1}(Upper-panel:) Left-figure represents that
the parameter $\omega$ over the charged lepton phase
$\varphi_{B}$. Right-figure represents the relation between the
mixing angle $\theta_{23}$ and the charged-lepton phase
$\varphi_{B}$. Here the horizontal dotted lines represent the
experimental lower and upper bounds of the mixing angle
$\theta_{23}$. (Lower-panel:) Left-figure shows the mixing angle
$\theta_{23}$ as a function of the parameter $\omega$. Here the
horizontal dotted lines represent the experimental upper and lower
bound of the mixing angle $\theta_{23}$. Right-figure shows the
mixing angle $\theta_{12}$ as a function of the parameter
$\omega$.}
\end{figure}

 In Fig.~\ref{Fig1},
the two figures of upper panel exhibit how the parameter
$\omega$ (left-panel) and the mixing angle $\theta_{23}$ (right-panel) are related with  the phase $\varphi_{B}$
for the case of $\phi_{A,B}=0$  at the GUT scale.
In this case we find that the parameters $\kappa$ and $\omega$
strongly depend on the phase $\varphi_{B}$, not
$\varphi_{A}$. The two figures of lower panel present the
predictions of $\theta_{23}$ (left-panel) and $\theta_{12}$ (right-panel) in terms of $\omega$.
The horizontal lines correspond to the bounds of present experimental values for $\theta_{23}$ and $\theta_{12}$
 given at $3\sigma$ range, Eq. (\ref{exp bound}), respectively.
{}From the results, it is interesting to see that most predictions of $\theta_{23}$ lie below $45^{\circ}$.
In fact, the experimental result for $\theta_{12}$ gives at 3$\sigma$ constraint the values of parameter
$0.4 \lesssim \omega \lesssim 1.05$.
We find that the constraint of $\omega$ prevents the prediction of $\theta_{23}$ from lying above $45^{\circ}$.

%%%%%%%%%%%%%%%
%    Fig 1    %
%%%%%%%%%%%%%%%

\begin{figure}[b]
%\vspace*{-5.0cm}
\hspace*{-2cm}
\begin{minipage}[t]{6.0cm}
\epsfig{figure=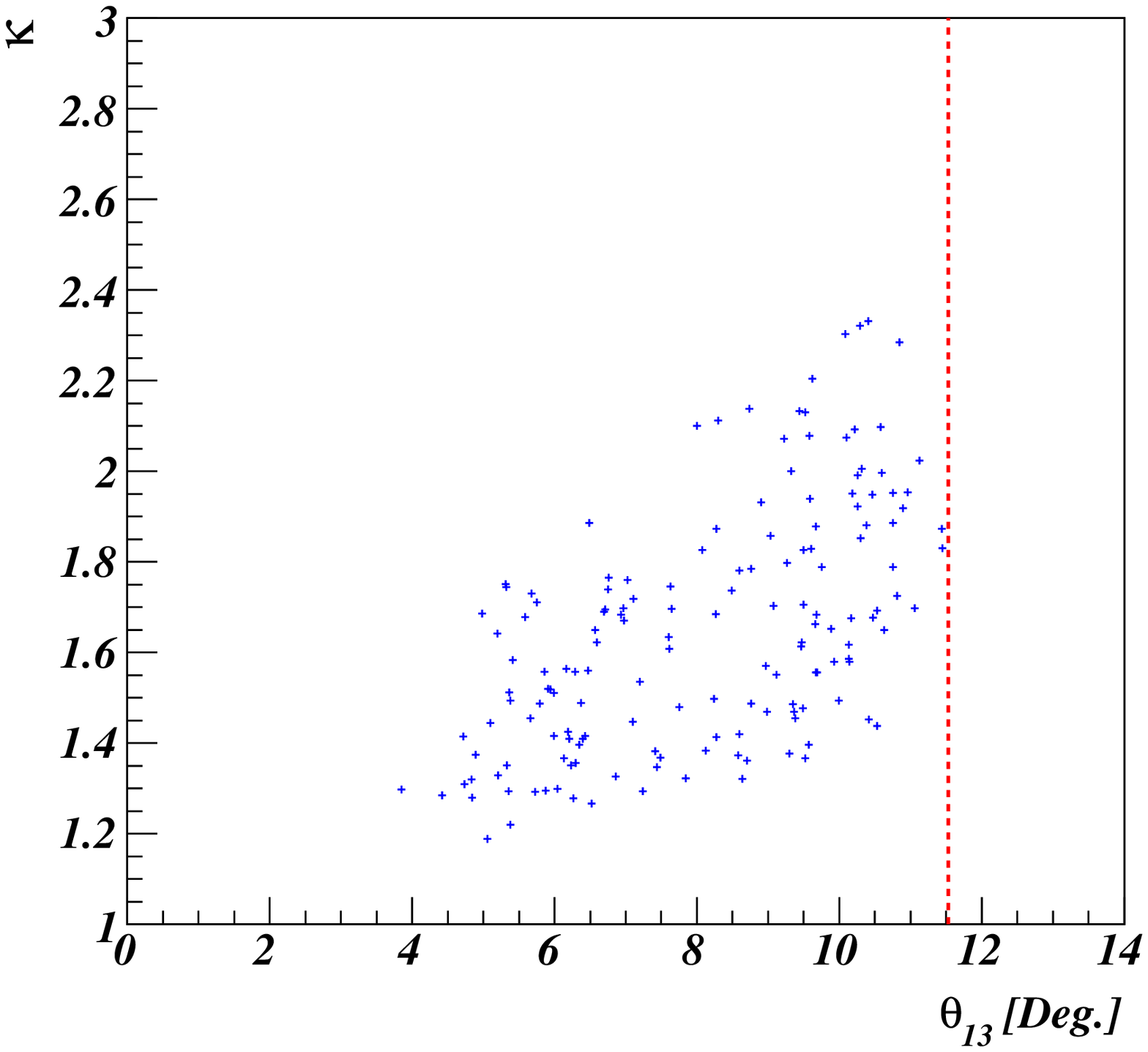,width=6.5cm,angle=0}
\end{minipage}
\hspace*{2.0cm}
\begin{minipage}[t]{6.0cm}
\epsfig{figure=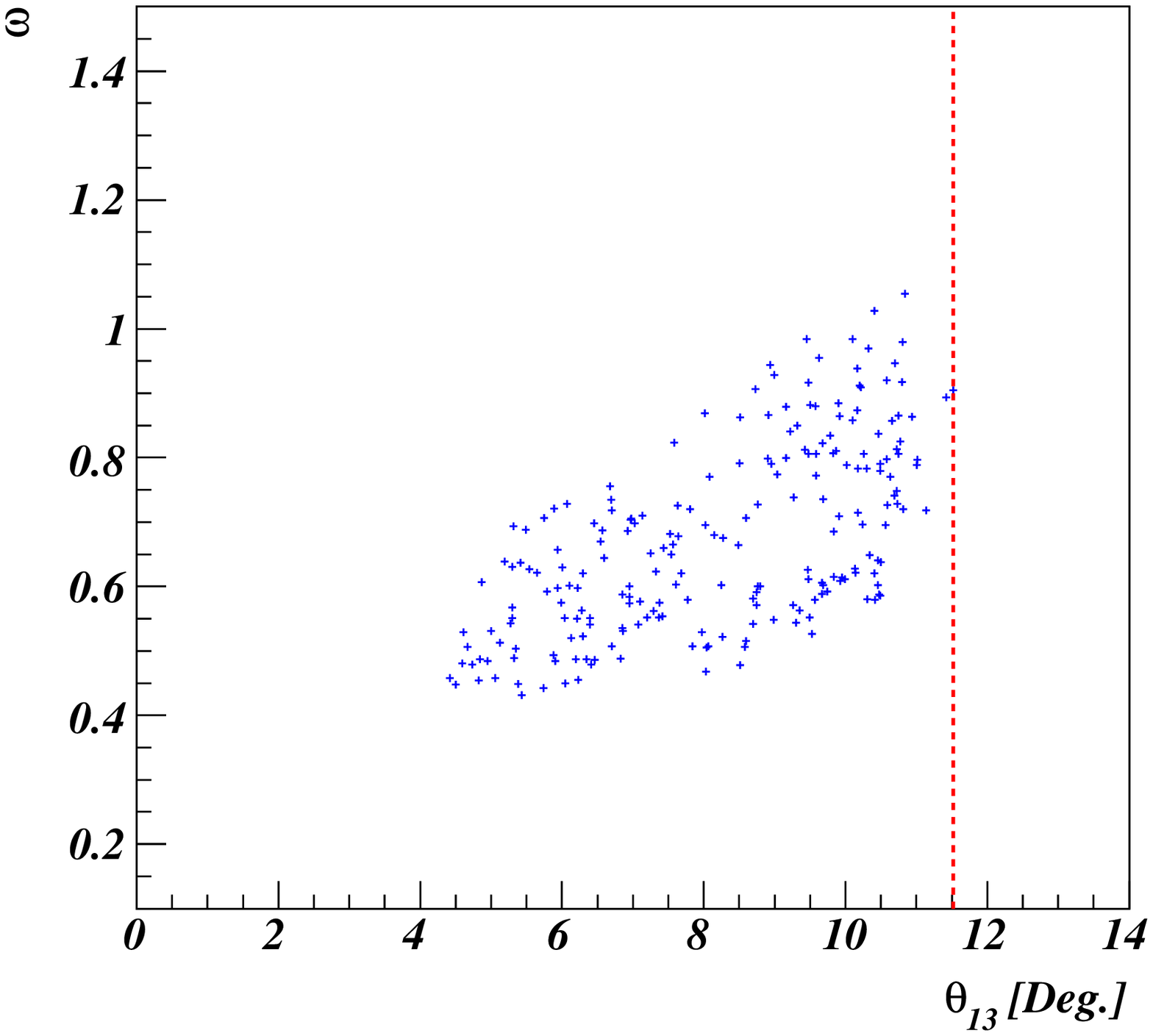,width=6.5cm,angle=0}
\end{minipage}
\vspace*{-1.0cm} \hspace*{-2cm}
\begin{minipage}[t]{6.0cm}
\epsfig{figure=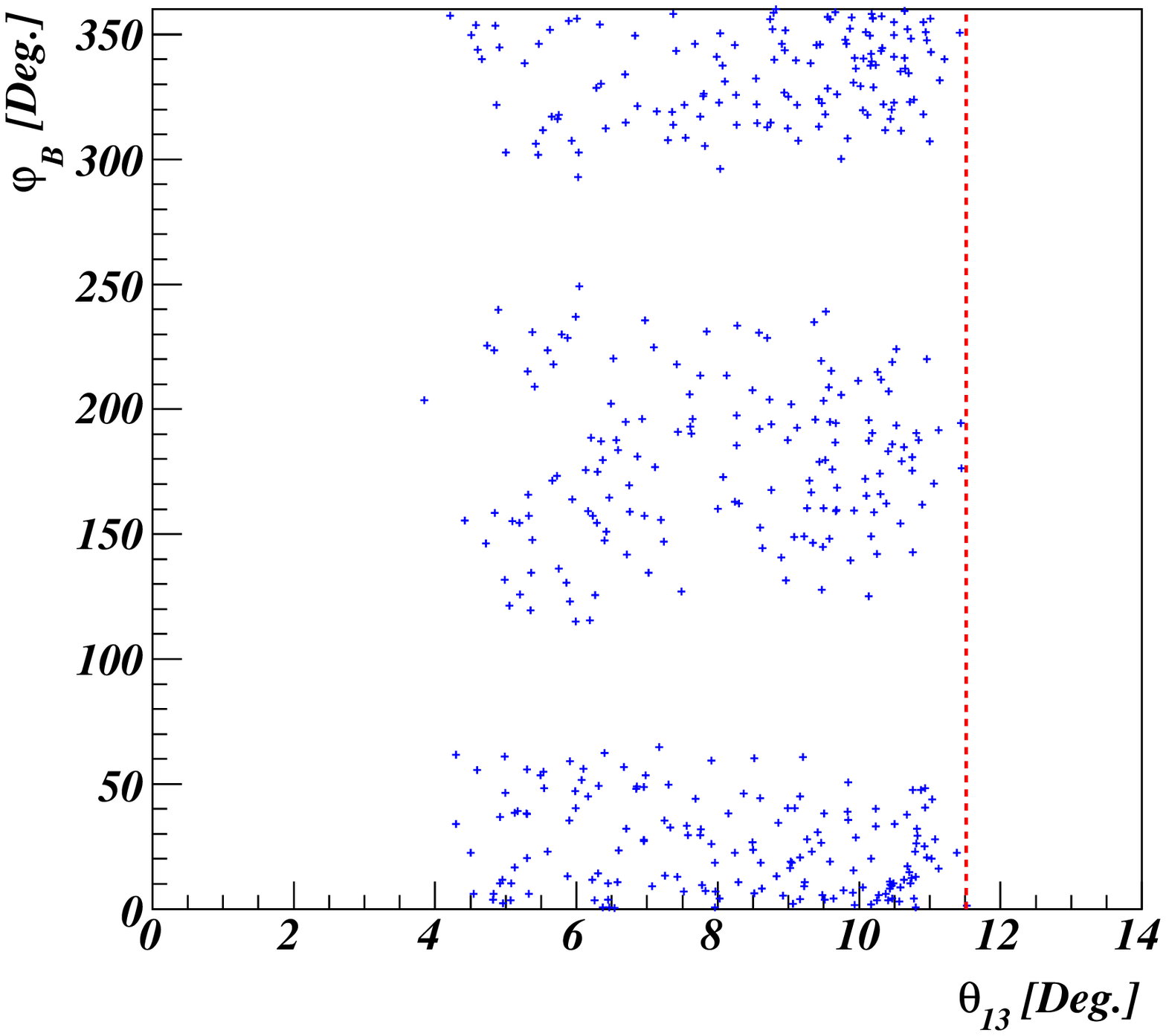,width=6.5cm,angle=0}
\end{minipage}
\hspace*{2.0cm}
\begin{minipage}[t]{6.0cm}
\epsfig{figure=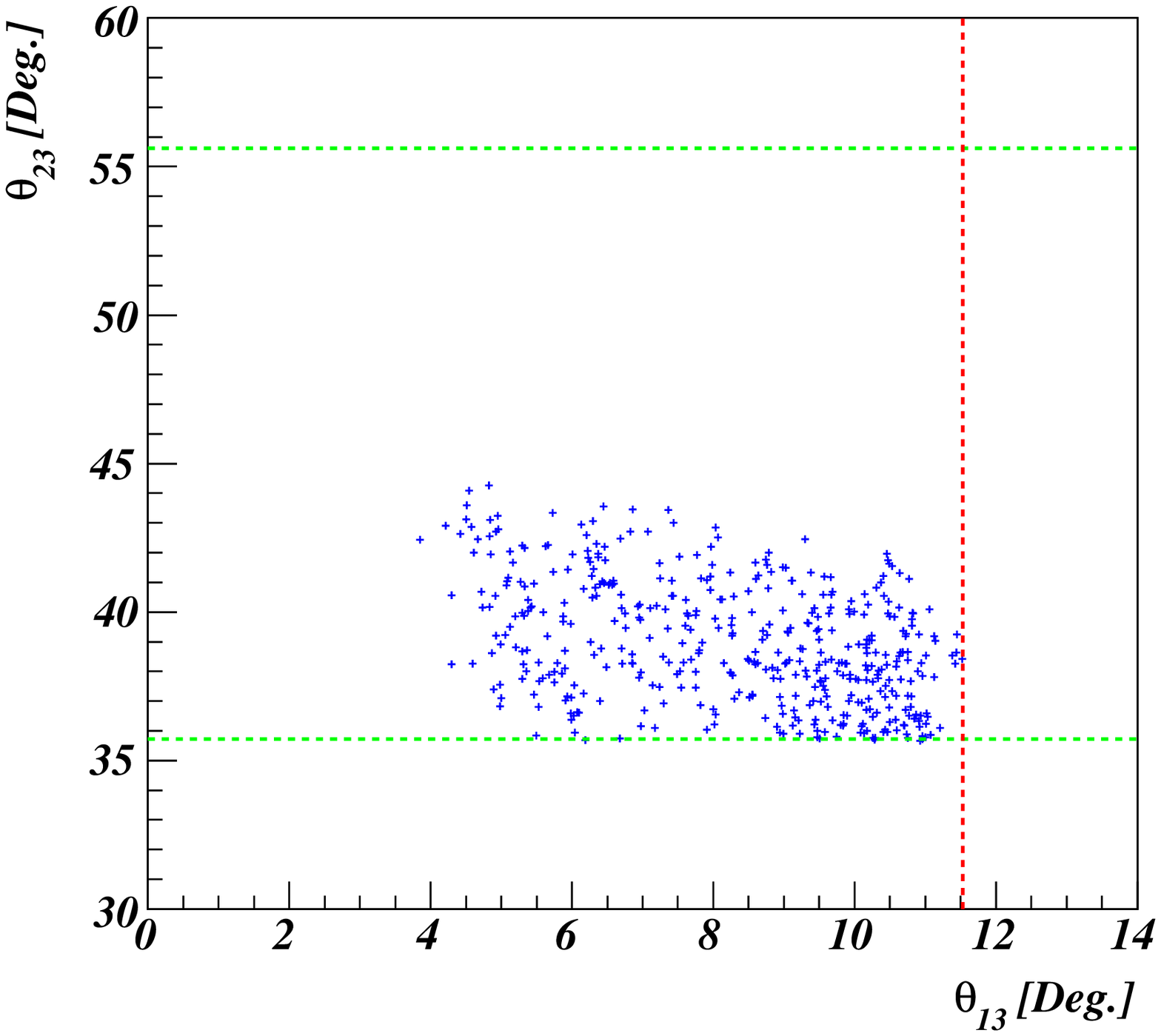,width=6.5cm,angle=0}
\end{minipage}
%\vspace*{-1.0cm}
\caption{\label{Fig2} In the case of $\phi_{A,B}=0,
\varphi_{A,B}\neq0$ at the GUT scale, the parameter regions
allowed by the $3\sigma$ experimental constraints for
$10^{6}\lesssim M[{\rm GeV}]\lesssim10^{12}$. (Upper-panel:)
Left-figure represents that the parameter $\kappa$ over the mixing
angle $\theta_{23}$ and right-figure $\omega$ over $\theta_{23}$,
where the vertical dotted line indicates the upper bound of
$\theta_{13}$. (Lower-panel:) Left-figure shows the charged-lepton
phase $\varphi_{B}$ over the mixing angle $\theta_{13}$, and the
vertical line corresponds to the upper bound on $\theta_{13}$.
Right-figure shows the predicted parameter space for $\theta_{13}$
and $\theta_{23}$ in FTY model and the horizontal dotted lines
indicate the experimental upper bound on $\theta_{13}$ and the
vertical dotted line represents the experimental lower and upper
bound on $\theta_{23}$.}
\end{figure}

Fig.~\ref{Fig2} shows how the mixing angle $\theta_{13}$  is predicted in terms of
the parameters $\kappa$, $\omega$ (upper-panel) and $\varphi_B$ (lower-panel) whose
sizes are constrained, as in Fig. 1, by the experimental results of $\theta_{23}$ and $\theta_{12}$.
In each figures, we draw the current reactor experimental upper bound on $\theta_{13}$.
We see from Fig.~\ref{Fig2} that very small values of $\theta_{13}$ are not predicted
in FTY model.
In lower right panel, we present the predicted regions for $\theta_{13}$ and $\theta_{23}$
in FTY model.

%%%%%%%%%%%%%%%
%    Fig 2    %
%%%%%%%%%%%%%%%

\begin{figure}[b]
%\vspace*{-5.0cm}
\hspace*{-2cm}
\begin{minipage}[t]{6.0cm}
\epsfig{figure=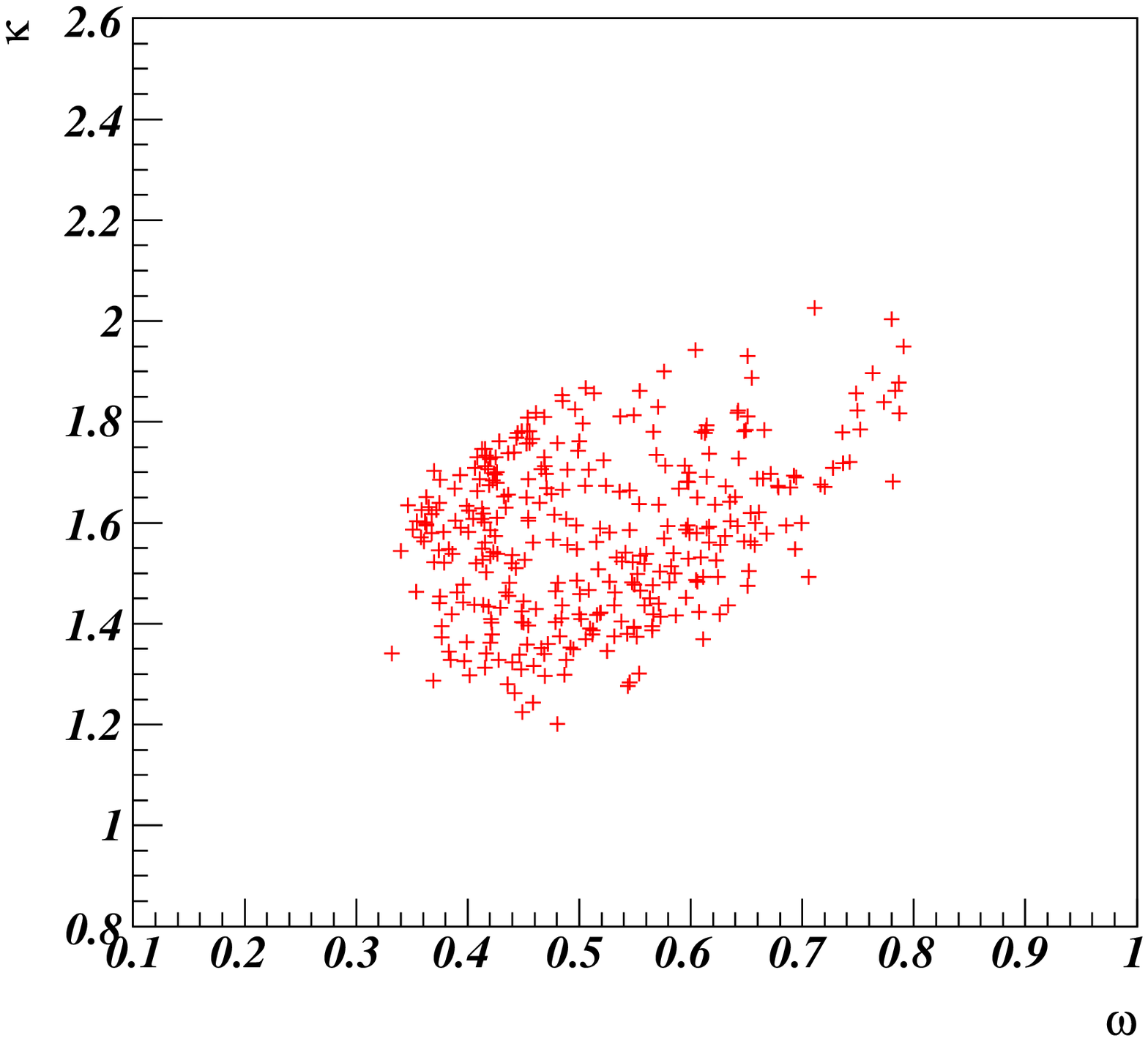,width=6.5cm,angle=0}
\end{minipage}
\hspace*{2.0cm}
\begin{minipage}[t]{6.0cm}
\epsfig{figure=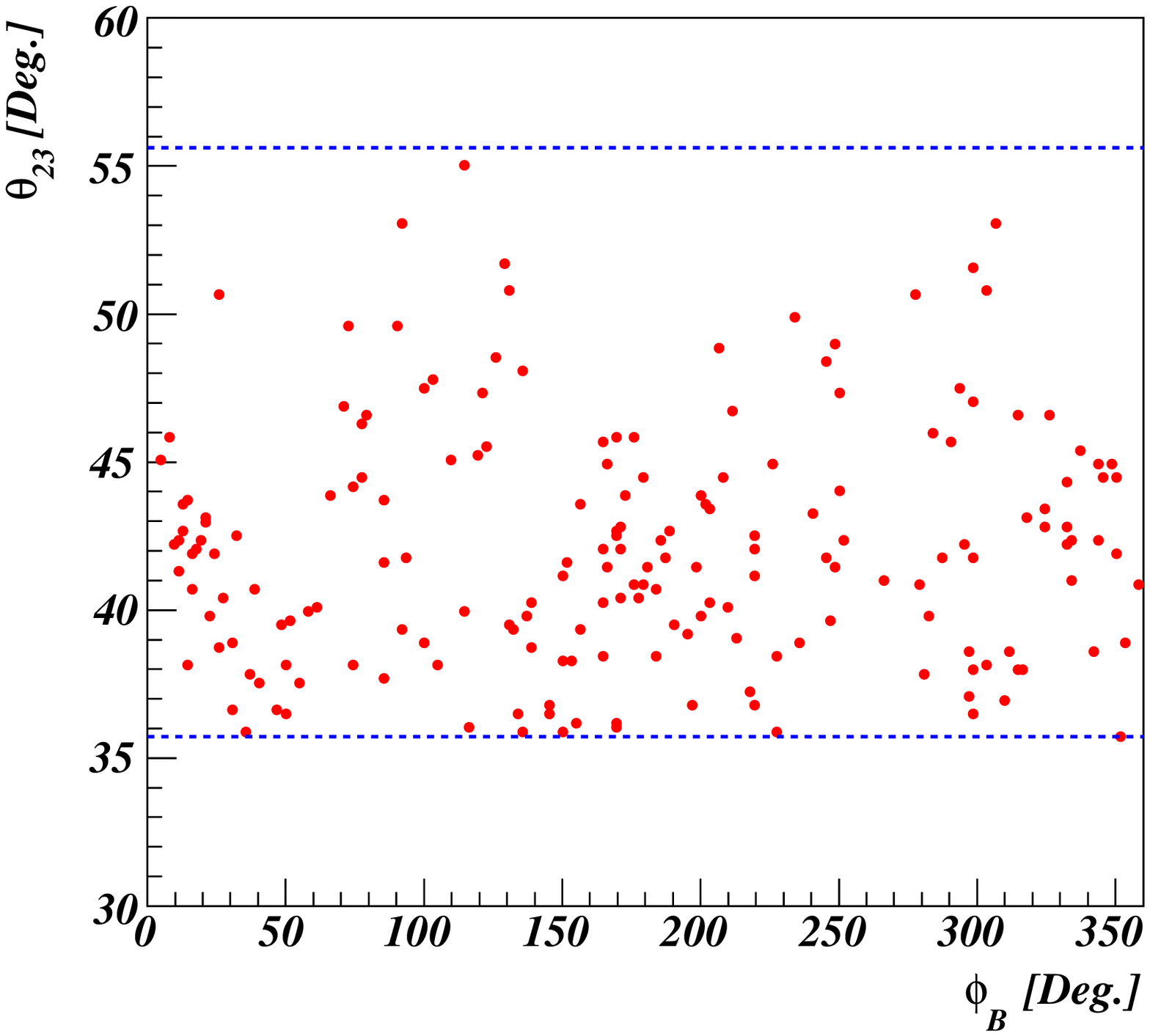,width=6.5cm,angle=0}
\end{minipage}
\vspace*{-1.0cm} \hspace*{-2cm}
\begin{minipage}[t]{6.0cm}
\epsfig{figure=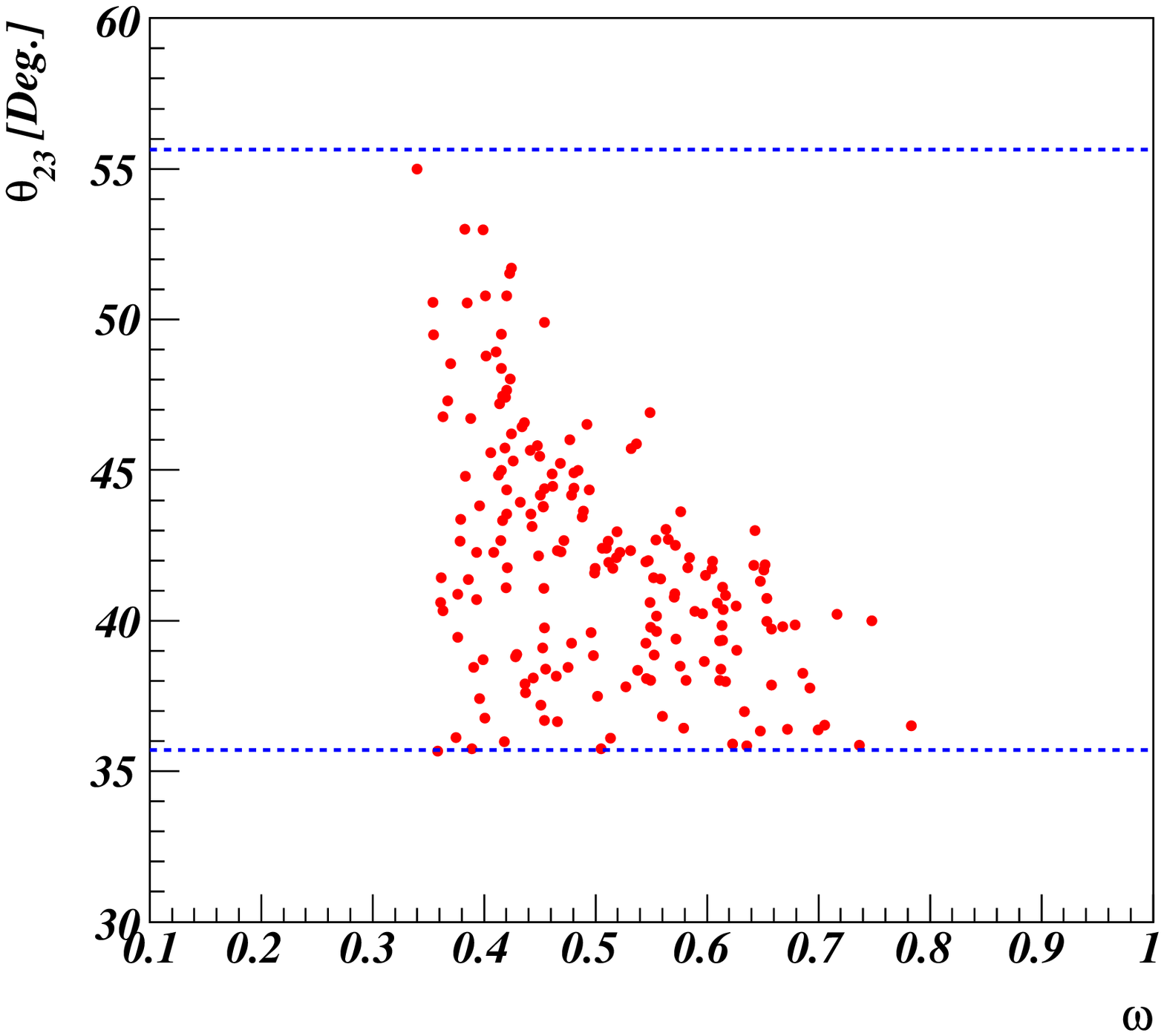,width=6.5cm,angle=0}
\end{minipage}
\hspace*{2.0cm}
\begin{minipage}[t]{6.0cm}
\epsfig{figure=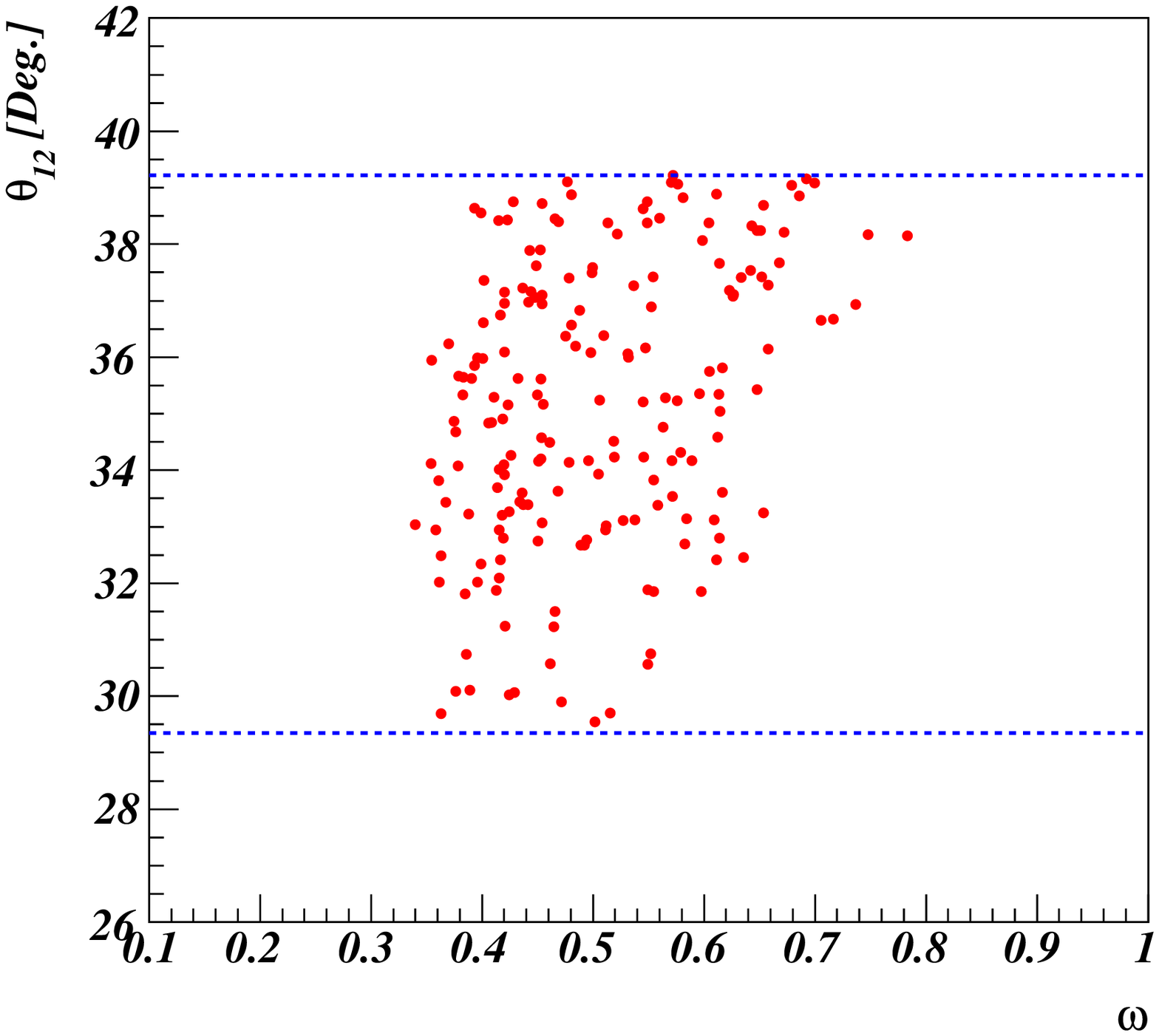,width=6.5cm,angle=0}
\end{minipage}
%\vspace*{-1.0cm}
\caption{\label{Fig3} (Upper-panel:) Left-figure represents the
allowed parameter space, $\kappa$ {\it vs.} $\omega$. Right-figure
represents the mixing angle $\theta_{23}$ as a function of
$\phi_B$. (Lower-panel:) Left-figure shows how the mixing angle
$\theta_{23}$ predicted in terms of $\omega$.  Right-figure shows
how $\theta_{12}$  predicted in terms of $\omega$. Here the
horizontal dotted lines represent the experimental upper and lower
bound of the mixing angle $\theta_{23}$ and $\theta_{12}$,
respectively.}
\end{figure}

Fig.~\ref{Fig3} shows the parameter spaces allowed by the
$3\sigma$ experimental constraints given in Eq. (\ref{exp bound})
for $10^{6}\lesssim M[{\rm GeV}]\lesssim10^{12}$ when the CP
phases $\phi_{A}$ and $\phi_{B}$ are turned on at the GUT scale.
The upper left panel plots the correlation between $\kappa$ and
$\omega$, and the upper right panel presents the predictions of
$\theta_{23}$ in terms of $\phi_B$. The lower left (right) panel
shows the prediction of $\theta_{23} (\theta_{12})$ in terms of
$\omega$. Contrary to the previous case with vanishing CP phases
$\phi_{A,B}$, the values above $45^{\circ}$ for $\theta_{23}$ are
possibly predicted.

%%%%%%%%%%%%%%%
%    Fig 3    %
%%%%%%%%%%%%%%%

\begin{figure}[b]
%\vspace*{-5.0cm}
\hspace*{-2cm}
\begin{minipage}[t]{6.0cm}
\epsfig{figure=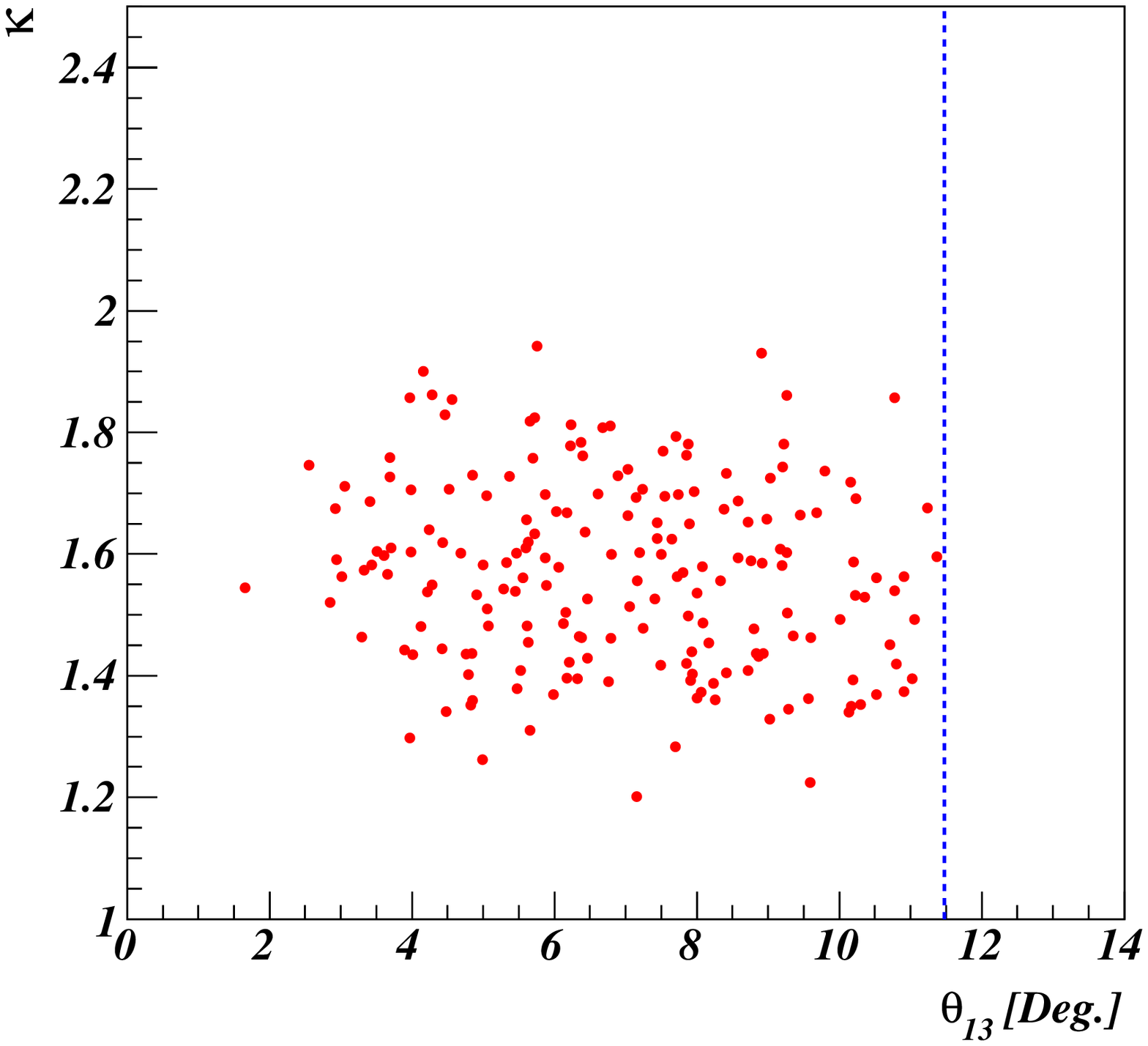,width=6.5cm,angle=0}
\end{minipage}
\hspace*{2.0cm}
\begin{minipage}[t]{6.0cm}
\epsfig{figure=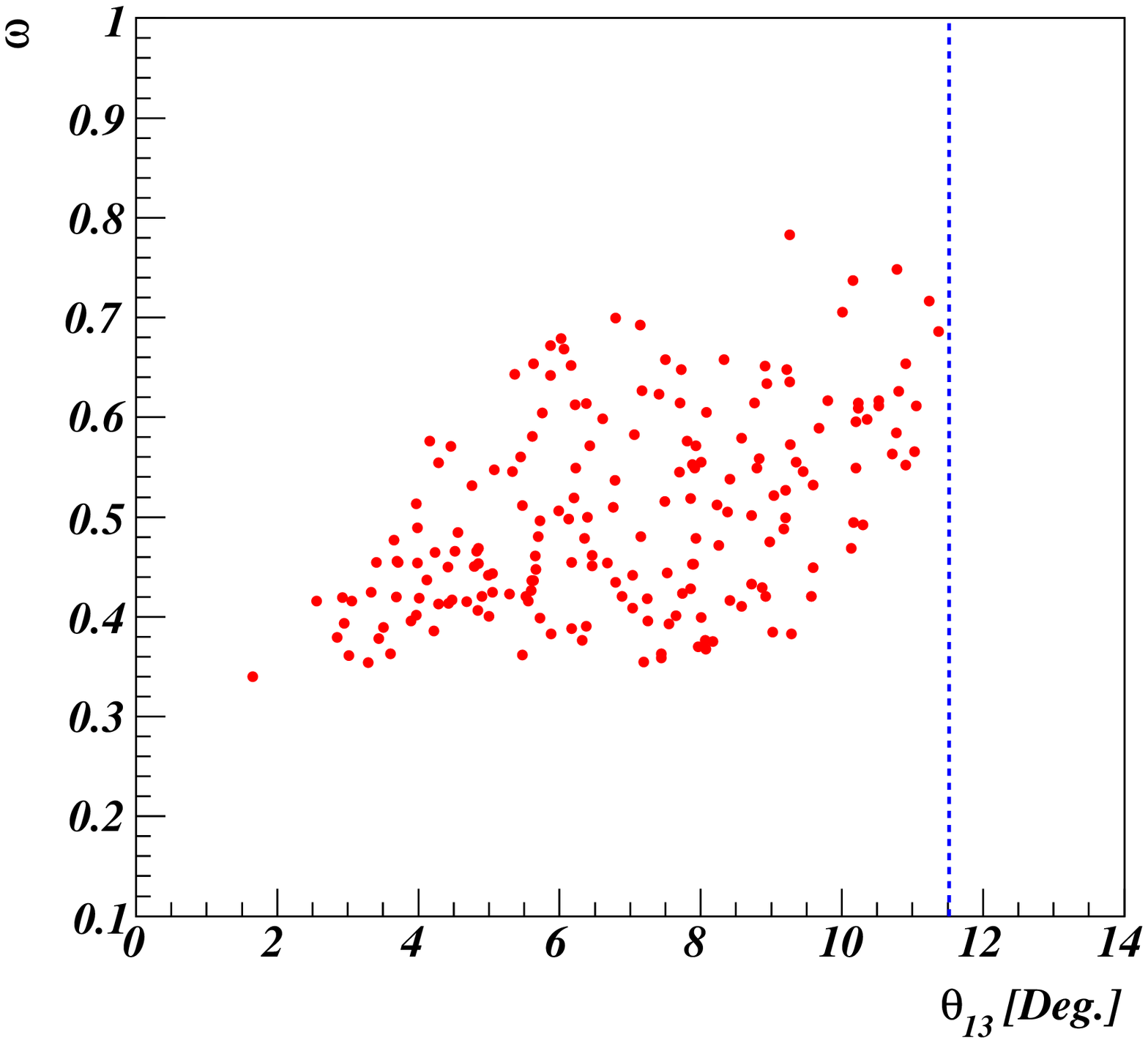,width=6.5cm,angle=0}
\end{minipage}
\vspace*{-1.0cm} \hspace*{-2cm}
\begin{minipage}[t]{6.0cm}
\epsfig{figure=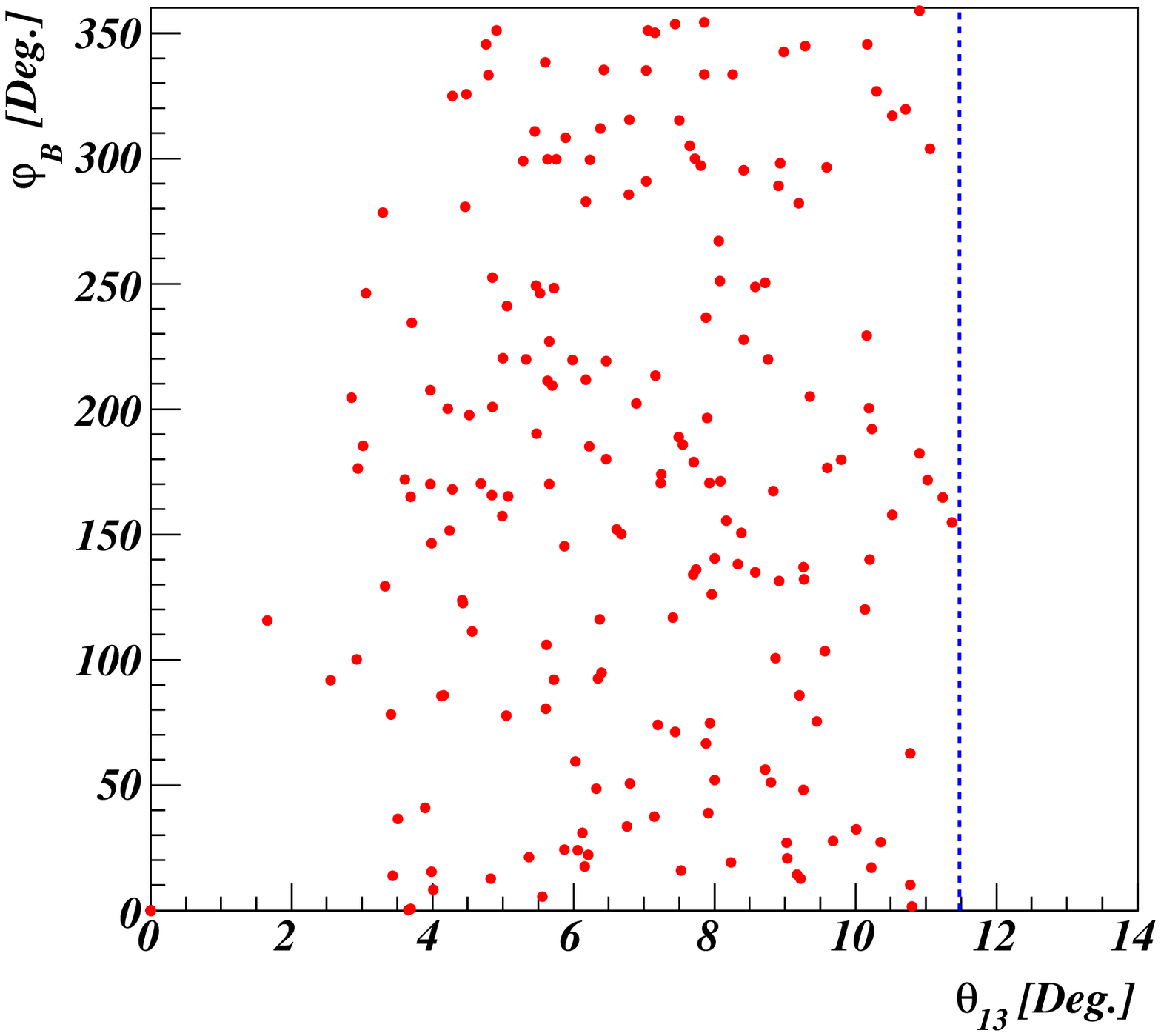,width=6.5cm,angle=0}
\end{minipage}
\hspace*{2.0cm}
\begin{minipage}[t]{6.0cm}
\epsfig{figure=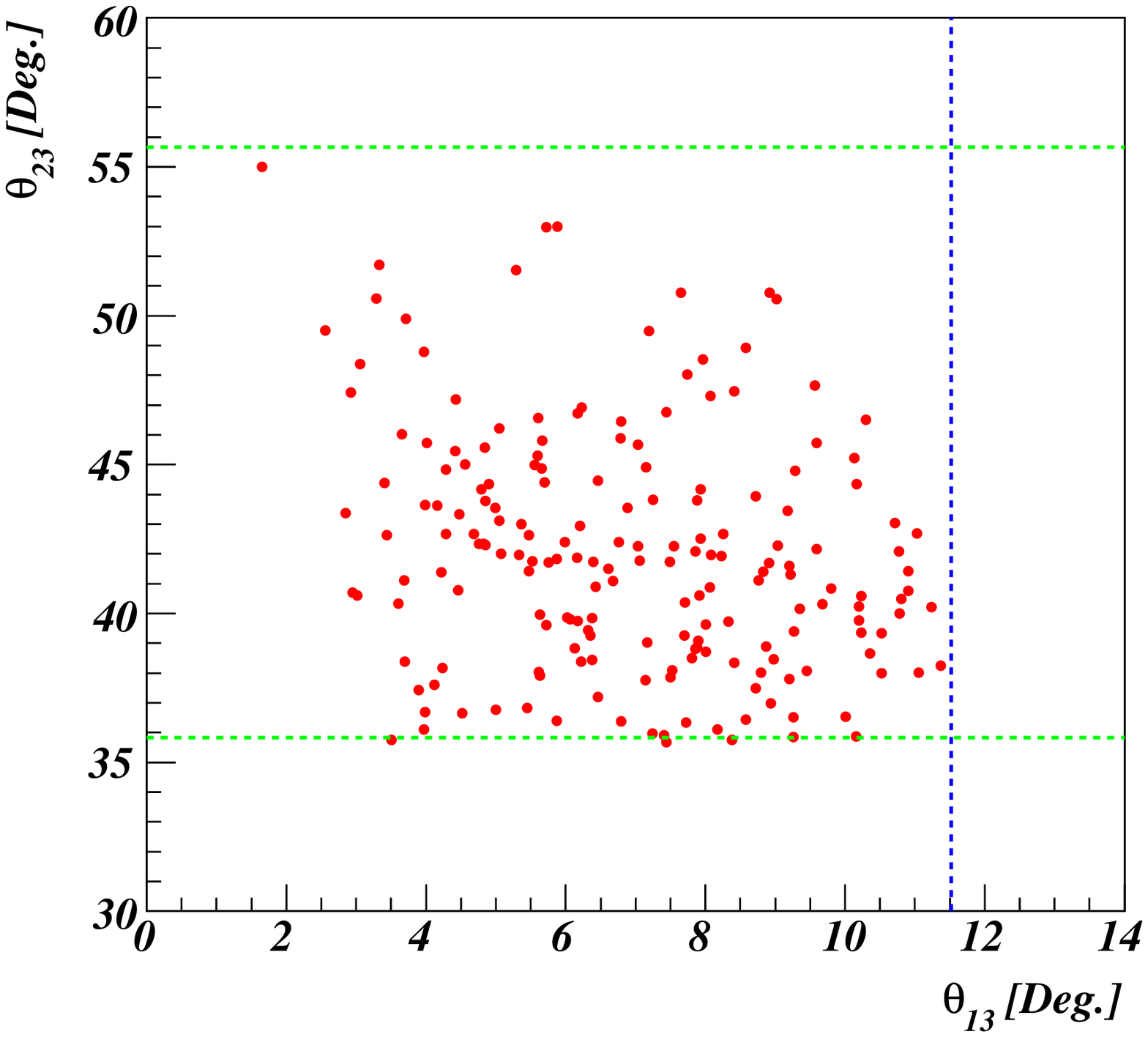,width=6.5cm,angle=0}
\end{minipage}
%\vspace*{-1.0cm}
\caption{\label{Fig4}  The same as Fig. \ref{Fig2} except for $\phi_{A,B}\neq 0$
at the GUT scale.}
\end{figure}

Similar to Fig.~\ref{Fig2}, we present in Fig.~\ref{Fig4} how the mixing angle $\theta_{13}$
is predicted in terms of
the parameters $\kappa$, $\omega$ (upper-panel) and $\varphi_B$ (down-panel), whose
sizes are constrained, as in Fig. 3, by the experimental results of $\theta_{23}$ and $\theta_{12}$.
In each figures, we draw the current reactor experimental upper bound on $\theta_{13}$.
We see from Fig.~\ref{Fig4} that very small values of $\theta_{13}$ are allowed
in FTY model, which is contrary to the previous case with $\phi_{A,B}=0$.
In lower right panel, we present the predicted regions for $\theta_{13}$ and $\theta_{23}$
in FTY model.

%%%%%%%%%%%%%%%
%    Fig 4    %
%%%%%%%%%%%%%%%

%%%%%%%%%%%%%%%%%%%%%%%%%%%%%%%%%%%%%%%%%%%%%%%%%%%%%%%%%%%%%%%%%%%%%%%%%%%%%%%%%%%%%%%%%%%%%%%%%%%%%%%%%%%%%%%%%%%%%%%%%%%%%%%%%%%%%%%%%%%%%%%%%%%%%%%%%%%%%%%%%%%%%%%%%%%%%%%%%%%%%%%%%%%%%%%%%%%%%%%%%%%%%%%%%%%%%%%%%%%%%%%%%%
\section{Radiatively induced resonant leptogenesis}

 It is well known
that if heavy Majorana neutrinos are exact degenerate as in FTY model, the
generated lepton asymmetry is zero \cite{liu}. A non-zero leptonic
asymmetry can be generated if and only if the $CP$-odd invariant
$\Delta_{\rm CP}={\rm
Im~Tr}[\textbf{Y}_{\nu}\textbf{Y}^{\dag}_{\nu}\textbf{M}_{R}\textbf{M}^{\dag}_{R}
\textbf{M}_{R}\textbf{Y}^{\ast}_{\nu}\textbf{Y}^{T}_{\nu}\textbf{M}^{\dag}_{R}]$
does not vanish \cite{Pilaftsis:2003gt}. The exact mass degeneracy
of three right-handed neutrinos implies that the $CP$-odd
invariant
 \begin{eqnarray}
  \Delta_{\rm CP}=2\sum_{i<j}\Big\{M_{i}M_{j}(M^{2}_{j}-M^{2}_{i}){\rm Im}[H_{ij}]{\rm Re}[H_{ij}]\Big\},~~~H\equiv
  \textbf{Y}_{\nu}\textbf{Y}^{\dag}_{\nu},
  \label{CP-invariant}
 \end{eqnarray}
which is relevant for leptogenesis \cite{Pilaftsis:1997jf}, is
actually vanishing. In order to accommodate leptogenesis, it
requires not only $M_{i}\neq M_{j}$ but also ${\rm Im}[H_{ij}]{\rm
Re}[H_{ij}]\neq0$. Even if we have exactly degenerate heavy Majorana
neutrinos at a certain high energy scale, it is likely that some
splitting in the mass spectrum could be induced at a different scale
through RG running effect. If this is the case, we will get the
splittings of heavy Majorana neutrino masses $i.e.$ a slightly
broken $SO(3)$ symmetry in the right-hand sector with $|M_{1}|\simeq
|M_{2}|\simeq |M_{3}|$. And the Dirac neutrino Yukawa matrix
$\textbf{Y}_{\nu}$ is also modified by the same RG effect, which is
very important to get non-zero ${\rm Im}[H_{ij}]{\rm
Re}[H_{ij}]\neq0$, as will be shown later.

 Let us consider the evolution of the right-handed heavy Majorana neutrinos masses
and the matrix $\Omega$ which diagonalizes the heavy Majorana mass matrix $\textbf{M}$ in lagrangian
(\ref{lagrangian}), whose RGEs can be written by \cite{RG1}\footnote{Actually, Ref. \cite{RG1} follows
bottom-up approach, that is, from electroweak scale to seesaw
scale. On the contrary, we apply  top-down approach.}:
 \begin{eqnarray}
  \frac{d}{dt}\textbf{M} &=& (\textbf{Y}'_{\nu}\textbf{Y}'^{\dag}_{\nu})\textbf{M}+\textbf{M}(\textbf{Y}'_{\nu}\textbf{Y}'^{\dag}_{\nu})^{T},\\
  \frac{d}{dt}\Omega &=& \Omega A,
  \label{RG 0}
 \end{eqnarray}
 where $t=\frac{1}{16\pi^{2}}ln(\mu/\Lambda)$ with renormalizable scale
$\mu$ and degenerate seesaw scale $\Lambda$ and
$\textbf{Y}'_{\nu}$ is the re-basing form in Eq. (\ref{rebasing}).
With the use of unitary transformation $N_{j}\rightarrow
\Omega_{ji}N_{i}$, one can obtain
 \begin{eqnarray}
  \Omega^{T}\textbf{M}\Omega=diag(M_{1}, M_{2}, M_{3}).
  \label{RG 1}
 \end{eqnarray}
 Since $(d/dt)\Omega=\Omega A$, $A$ satisfies
$A+A^{\dag}=0$, and then from Eq. (\ref{RG 1}) we can obtain:
 \begin{eqnarray}
  \frac{dM_{i}\delta_{ij}}{dt}=A^{T}_{ij}M_{j}+M_{i}A_{ij}+\{\Omega^{T}[(\textbf{Y}'_{\nu}\textbf{Y}'^{\dag}_{\nu})\textbf{M}
             +\textbf{M}(\textbf{Y}'_{\nu}\textbf{Y}'^{\dag}_{\nu})^{T}]\Omega\}_{ij}.
 \label{RG 2}
 \end{eqnarray}
Thus, the RG evolutions for  the right-handed heavy Majorana neutrino masses are governed by the diagonal part in the above equation:
 \begin{eqnarray}
  \frac{dM_{i}}{dt}=2M_{i}(Y_{\nu}Y^{\dag}_{\nu})_{ii},~~~\text{with}~Y_{\nu}=\Omega^{T}\textbf{Y}'_{\nu}
  \label{RG 3}
 \end{eqnarray}
and the anti-hermitian property of the imaginary part of the matrix $A$ leads to
 \begin{eqnarray}
  {\rm Im}[A_{ii}]=0.
  \label{RG 4}
 \end{eqnarray}
 In addition, the off-diagonal part in Eq. (\ref{RG 2}) leads to
 \begin{eqnarray}
  A_{jk}=\frac{M_{k}+M_{j}}{M_{k}-M_{j}}{\rm Re}[(Y_{\nu}Y^{\dag}_{\nu})_{jk}]+i\frac{M_{j}-M_{k}}{M_{j}+M_{k}}{\rm Im}[(Y_{\nu}Y^{\dag}_{\nu})_{jk}]=-A^{\ast}_{kj}, ~(j\neq k).
  \label{RG 5}
 \end{eqnarray}

The RG equation for the Dirac neutrino Yukawa matrix  is given by
 \begin{eqnarray}
   \frac{dY_{\nu}}{dt} &=&
   Y_{\nu}[(T-\frac{3}{4}g^{2}_{Y}-\frac{9}{4}g^{2}_{2})-\frac{3}{2}(\textbf{Y}^{\dag}_{l}\textbf{Y}_{l}
   -Y^{\dag}_{\nu}Y_{\nu})]+A^{T}Y_{\nu},
  \label{RG 6}
 \end{eqnarray}
where
$T=Tr(3\textbf{Y}^{\dag}_{u}\textbf{Y}_{u}+3\textbf{Y}^{\dag}_{d}\textbf{Y}_{d}
  +Y^{\dag}_{\nu}Y_{\nu}+\textbf{Y}^{\dag}_{l}\textbf{Y}_{l})$, $\textbf{Y}_{u}~(\textbf{Y}_d)$
  and $\textbf{Y}_{l}$ are
the Yukawa matrices for up-type (down-type) quarks and charged
leptons and $g_{2,Y}$ are the $SU(2)_{L}$ and $U(1)_{Y}$ gauge
coupling constants. The RG evolution for the quantity $H$ relevant
for leptogenesis can be written as
 \begin{eqnarray}
%  \frac{d}{dt}(Y_{\nu}Y^{\dag}_{\nu})
%    = 2Y_{\nu}\{Q+P_{\nu}\}Y^{\dag}_{\nu}+A^{T}Y_{\nu}Y^{\dag}_{\nu}+Y_{\nu}Y^{\dag}_{\nu}A^{\ast},
\frac{d}{dt}H
       = 2Y_{\nu}\{Q+P_{\nu}\}Y^{\dag}_{\nu}+A^{T}H+HA^{\ast},
  \label{RG 7}
 \end{eqnarray}
where
 \begin{eqnarray}
 Q= T-\frac{3}{4}g^{2}_{2}-\frac{9}{4}g^{2}_{1},~~~~~~
 P_{\nu}=-\frac{3}{2}(\textbf{Y}^{\dag}_{l}\textbf{Y}_{l}-Y^{\dag}_{\nu}Y_{\nu}).\nonumber
 \end{eqnarray}

{}From Eq. (\ref{RG 4}), we see that there exists a singularity in
$A_{jk}$. The singularity in $A_{jk}$ can be eliminated with the
help of an appropriate rotation between degenerate heavy Majorana
neutrino states. Such a rotation does not change any physics and
it is equivalent to absorb the rotation matrix $R$ into the
neutrino Dirac Yukawa matrix $Y_\nu$,
\begin{eqnarray}
  Y_{\nu}\rightarrow \widetilde{Y}_{\nu}=RY_{\nu},
  \label{transform}
 \end{eqnarray}
where the matrix $R$ matrix is an $3\times 3$ orthogonal matrix
which can be parameterized in terms of angles $\theta_{i}$ as
$R(\theta_{i},\theta_{j},\theta_{k})=R(\theta_{i})\cdot
R(\theta_{j})\cdot R(\theta_{k})$
\begin{eqnarray}
R(\theta_{1})=\left(
\begin{array}{ccc}
 1 & 0 & 0 \\
 0 & c_{1} & s_{1} \\
 0 & -s_{1} & c_{1} \\
\end{array}
\right),~R(\theta_{2})=\left(
\begin{array}{ccc}
 c_{2} & 0 & s_{2} \\
 0 & 1 & 0 \\
 -s_{2} & 0 & c_{2} \\
\end{array}
\right),~R(\theta_{3})=\left(
\begin{array}{ccc}
 c_{3} & s_{3} & 0 \\
 -s_{3} & c_{3} & 0 \\
 0 & 0 & 1 \\
\end{array}
\right),\label{R1}
\end{eqnarray}
where $s_{i}\equiv\sin\theta_{i}, c_{i}\equiv\cos\theta_{i}$.
Then, the singularity in real part of $A_{jk}$ can be indeed
removed when the rotation angles $\theta_{i}$ are taken to be
satisfied with the condition,
 \begin{eqnarray}
 {\rm Re}[(\widetilde{Y}_{\nu}\widetilde{Y}^{\dag}_{\nu})_{jk}]=0~~~\text{for any pair $j,k$
  corresponding to}~M_{j}=M_{k}.
 \label{singularity2}
 \end{eqnarray}

 At the
degeneracy scale of $\textbf{M}_{R}$ there is  a freedom to rotate the
right-handed neutrino fields $N_{1,2,3}$ with a real orthogonal
matrix that does not change $\textbf{M}_{R}$, but rotates $Y_{\nu}$ to the
appropriate basis, which allows the use of an $SO(3)$ transformation
to remove the off-diagonal elements of ${\rm Re}[H]$, and thus we
can obtain a matrix $\widetilde{H}$ satisfying the condition Eq.
(\ref{singularity2}) as follows,
 \begin{equation}
  \widetilde{H}\equiv(\widetilde{Y}_{\nu}\widetilde{Y}^{\dag}_{\nu})=RHR^{T}
  =B_{\nu}^{2}\left(\begin{array}{ccc}
  \widetilde{h}_{11}   &  i{\rm Im}[\widetilde{h}_{12}]  & i{\rm Im}[\widetilde{h}_{13}]  \\
  -i{\rm Im}[\widetilde{h}_{12}]  &  \widetilde{h}_{22}   & i{\rm Im}[\widetilde{h}_{23}] \\
  -i{\rm Im}[\widetilde{h}_{13}] &  -i{\rm Im}[\widetilde{h}_{23}]  & \widetilde{h}_{33}
\end{array}
\right),
 \label{Htildeplus1}
 \end{equation}
 where $\widetilde{h}_{jj}$ and ${\rm Im}[\widetilde{h}_{jk}]$
$(j\neq k=1,2,3)$ are given by
  \begin{eqnarray}
  \widetilde{h}_{11} &=& \omega^{2}+\sin^{2}\theta_{3}+q_{1}\tan\theta_{2},\nonumber\\
  \widetilde{h}_{22} &=& (\omega^{2}+\cos^{2}\theta_{3})\cos^{2}\theta_{1}+\{1+\kappa^{2}-q_{1}\tan\theta_{2}\}\sin^{2}\theta_{1}+ q_{2}\frac{\sin2\theta_{1}}{\cos\theta_{2}},\nonumber\\
  \widetilde{h}_{33} &=& (\omega^{2}+\cos^{2}\theta_{3})\sin^{2}\theta_{1}+\{1+\kappa^{2}-q_{1}\tan\theta_{2}\}\cos^{2}\theta_{1}-
  q_{2}\frac{\sin2\theta_{1}}{\cos\theta_{2}},\nonumber\\
  {\rm Im}[\widetilde{h}_{12}] &=& \cos\theta_{3}\{\omega\sin(\phi_{A}-\phi_{B})\sin\theta_{1}-\kappa\sin\phi_{B}\cos\theta_{1}\sin\theta_{2}\}\nonumber\\
  &+&\sin\theta_{3}\{\kappa\sin\phi_{B}\sin\theta_{1}+\omega\sin(\phi_{A}-\phi_{B})\cos\theta_{1}\sin\theta_{2}\}, \nonumber\\
  {\rm Im}[\widetilde{h}_{13}] &=& \cos\theta_{3}\{\omega\sin(\phi_{A}-\phi_{B})\cos\theta_{1}+\kappa\sin\phi_{B}\sin\theta_{1}\sin\theta_{2}\}\nonumber\\
  &+&\sin\theta_{3}\{\kappa\sin\phi_{B}\cos\theta_{1}-\omega\sin(\phi_{A}-\phi_{B})\sin\theta_{1}\sin\theta_{2}\}, \nonumber\\
  {\rm Im}[\widetilde{h}_{23}] &=& \cos\theta_{2}\{\kappa\sin\phi_{B}\cos\theta_{3}-\omega\sin(\phi_{A}-\phi_{B})\sin\theta_{3}\}.
 \end{eqnarray}
Here, the parameters $q_1$ and $q_2$ are given by
  \begin{eqnarray}
  q_{1} &=& \kappa\cos\phi_{B}\sin\theta_{3}+\omega\cos(\phi_{A}-\phi_{B})\cos\theta_{3},\nonumber\\
  q_{2} &=& \kappa\cos\phi_{B}\cos\theta_{3}-\omega\cos(\phi_{A}-\phi_{B})\sin\theta_{3}.
 \label{Htildeplus2}
 \end{eqnarray}
 The angle $\theta_{i}$ in the real $3\times3$ orthogonal matrix $R$ and CP-violating parameters $\phi_{A}, \phi_{B}$
 in the matrix $Y_{\nu}$ make $d\widetilde{Y}/dt$
non-singular, $i.e.$ when the degeneracy is exact, $Y_{\nu}$ changes
rapidly from its unperturbed form at $t=0$  to a stable form that
makes $d\widetilde{Y}_{\nu}/dt$ non-singular in Eq. (\ref{RG 6}).
In the case of exact degenerate heavy Majorana neutrinos, $i.e.$,
$\textbf{M}_{R}=M\textbf{I}$,
% even though three of the real parameters of
%$R$ can be rotated away,
the rotation matrix $R$ must be used to remove singularities at the
degeneracy scale, therefore $\theta_{i}~ (i=1,2,3)$ is no longer
free parameters, $i.e.$, it is constrained by the conditions Eq.
(\ref{singularity2}) from which we can obtain the following
relations,
 \begin{eqnarray}
  \tan2\theta_{1} &=&
  \frac{2q_{2}}{\cos\theta_{2}(\omega^{2}+\cos^{2}\theta_{3}+q_{1}\tan\theta_{2}-1-\kappa^{2})},\nonumber\\
  \tan2\theta_{2} &=&
  \frac{2q_{1}}{\omega^{2}+\sin^{2}\theta_{3}-1-\kappa^{2}},~~
   \Big({\rm or}~\tan\theta_{2}=-\frac{\sin2\theta_{3}}{2q_{2}}\Big),
 \label{InfinityS}
 \end{eqnarray}
which show the initial  stable conditions
of angles at the GUT scale.  Note that $\theta_{1}, \theta_{2}$ and
$\theta_{3}$ have scale dependence when RG running from GUT to
seesaw scale,  Eq. (\ref{RG 0}).

%%%%%%%%%%%%%%%%%%%%%%%%%%%%%%%%%%%%%%%%%%%%%%%%%%%%%%%%%%%%%%%%%%%%%%%%%%%%%%%%%%%%%%%%%%%%%%%%%%%%%%%%%%%%%%%%%%%%%%%%%%%%%%%%%%%%%%%%%%%%%%%%%%%%%%%%%%%%%%%%%%%%%%%%%%%%%%%%%%%%%%%%%
\subsection{Flavor Independent Leptogenesis}

 In a basis where the right-handed Majorana neutrino
mass matrix is diagonal, ignoring flavor effects in the Boltzmann
evolution of charged leptons, the CP asymmetry generated through
the interference between tree and one-loop diagrams of heavy
singlet Majorana neutrino decay is given by \cite{lepto2,lepto}:
 \begin{eqnarray}
  \varepsilon_{i} = \frac{\sum_{\alpha}[\Gamma(N_{i}\rightarrow l_{\alpha}\varphi)
  -\Gamma(N_{i}\rightarrow \overline{l}_{\alpha}\varphi^{\dag})]}{\sum_{\alpha}[\Gamma(N_{i}\rightarrow l_{\alpha}\varphi)
  +\Gamma(N_{i}\rightarrow\overline{l}_{\alpha}\varphi^{\dag})]}
  = \frac{1}{8\pi(Y_{\nu}Y^{\dag}_{\nu})_{ii}}\sum_{j\neq i}{\rm
  Im}\Big\{(Y_{\nu}Y^{\dag}_{\nu})^{2}_{ij}\Big\}g\Big(\frac{M^{2}_{j}}{M^{2}_{i}}\Big),
 \label{cpasym0}
 \end{eqnarray}
where the function $g(x)$ is given by
 \begin{eqnarray}
  g(x)= \sqrt{x}\Big[\frac{1}{1-x}+1-(1+x){\rm ln}\frac{1+x}{x}\Big]
  \label{decayfunction}
  \end{eqnarray}
with $x=M^{2}_{j}/M^{2}_{i}$. In the case that the mass splitting of
the heavy Majorana neutrinos is very small, the CP asymmetries
$\varepsilon_{i}$ can be simplified by
%if all one-loop vertex corrections have been neglected
 \cite{lepto2, resonant} as
 \begin{eqnarray}
  \varepsilon_{i}\simeq \frac{{\rm Im}[(Y_{\nu}Y^{\dag}_{\nu})^{2}_{ij}]}{16\pi(Y_{\nu}Y^{\dag}_{\nu})_{ii}\delta^{ij}_{N}}
  \Big(1+\frac{\Gamma^{2}_{j}}{4M^{2}_{j}{\delta^{ij}_{N}}^{2}}\Big)^{-1},~~~
  {\rm with}~\Gamma_{j}=\frac{[Y_{\nu}Y^{\dag}_{\nu}]_{jj}M_{j}}{8\pi}~~(i\neq j=1,2,3),
   \label{eps}
 \end{eqnarray}
 where $j$ denotes a generation number and
$\Gamma_{j}$ is the decay width of the $j$th-generation right-handed neutrino.
We notice from Eq. (\ref{eps}) that $\varepsilon_{i}$ is resonantly enhanced when
$\Gamma_{j}\simeq(M^{2}_{i}-M^{2}_{j})/M_{i}$. Here, the parameter
$\delta^{jk}_{N}(=1-|M_{k}|/|M_{j}|\ll 1)$ reflecting the mass
splitting of the degenerate heavy Majorana neutrinos is
 governed by the following RGE  derived from Eq. (\ref{RG 2}),
           \begin{eqnarray}
            \frac{d\delta^{jk}_{N}}{dt}=
            2(1-\delta^{jk}_{N})[\widetilde{H}_{jj}-\widetilde{H}_{kk}],
            \label{deltaN}
           \end{eqnarray}
 which represents that radiative corrections induce mass-splittings proportional to
the neutrino couplings. In the limit $\delta^{jk}_{N}\ll1$, the
leading-log approximation for $\delta^{jk}_{N}$ can be easily found
to be
           \begin{eqnarray}
            \delta^{jk}_{N}\simeq2[\widetilde{H}_{jj}-\widetilde{H}_{kk}]\cdot
            t.
            \label{deltaN1}
           \end{eqnarray}
In order for Eq. (\ref{cpasym0}) to give successful leptogenesis,
not only the degeneracy of right-handed neutrinos
should be broken but also the non-vanishing ${\rm Im}[(Y_{\nu}Y^{\dag}_{\nu})^{2}_{ik}]$
is required at seesaw scale $M$.

To see how leptogenesis can successfully be achieved, let us first
consider the case that $\phi_{A}=\phi_{B}=0$ in $Y_{\nu}$ Eq.
(\ref{FTY1}) at the GUT scale, while keeping CP phases arisen from
the charged-lepton Yukawa matrix $\textbf{Y}_{l}$ which move to
$Y_{\nu}$ through re-basing, $i.e.$,
$\widetilde{Y}_{\nu}=RY'_{\nu}=RY_{\nu}V$. In this case, the
off-diagonal elements of
$\widetilde{H}\equiv(\widetilde{Y}_{\nu}\widetilde{Y}^{\dag}_{\nu})$
becomes zero, so that CP asymmetry could not be generated. However,
the RG effects mainly due to $Y_{\tau}$ lead to non-vanishing
off-diagonal elements in $\widetilde{H}_{jk}$, whose forms are
approximately given by
\begin{eqnarray}
 {\rm Re}[\widetilde{H}_{jk}(t)] &\simeq& -\frac{3}{2}y^{2}_{\tau}{\rm Re}[(\widetilde{Y}_{\nu})_{j3}(\widetilde{Y}^{\dag}_{\nu})_{3k}]
           \cdot t,\nonumber\\
 {\rm Im}[\widetilde{H}_{jk}(t)] &\simeq& -3y^{2}_{\tau}{\rm
 Im}[(\widetilde{Y}_{\nu})_{j3}(\widetilde{Y}^{\dag}_{\nu})_{3k}]\cdot t.
            \label{RGeffect21}
 \end{eqnarray}
{}From these results, we see that CP-violating effects are induced by RG corrections
due to the charged-lepton Yukawa
couplings, which can play a crucial role in leptogenesis
\cite{radiative}. With the help of Eqs. (\ref{cpasym0},\ref{RGeffect21}),
the CP-asymmetry for each heavy Majorana
neutrino is given as
 \begin{eqnarray}
  \varepsilon_{i} \simeq \frac{9y^{4}_{\tau}}{512\pi^{3}\widetilde{H}_{ii}}\cdot{\rm ln}\Big(\frac{M_{i}}{\Lambda}\Big)
  \sum_{j}\frac{{\rm Re}[(\widetilde{Y}_{\nu})_{j3}(\widetilde{Y}^{\dag}_{\nu})_{3i}]{\rm Im}[(\widetilde{Y}_{\nu})_{j3}(\widetilde{Y}^{\dag}_{\nu})_{3i}]}
  {\widetilde{H}_{jj}-\widetilde{H}_{ii}}.
  \label{RGLepto00}
 \end{eqnarray}

 %which can be classified in our work as
 %\begin{eqnarray}
 % {\rm Im}[(\widetilde{H}_{jk})^{2}] &\simeq& 9y^{4}_{\tau}{\rm Re}[(\widetilde{Y}_{\nu})_{j3}(\widetilde{Y}^{\dag}_{\nu})_{3k}]
 % {\rm Im}[(\widetilde{Y}_{\nu})_{j3}(\widetilde{Y}^{\dag}_{\nu})_{3k}]\cdot t^{2}~~~{\rm
 % for}~~\phi_{A}=\phi_{B}=0,\nonumber\\
 % {\rm Im}[(\widetilde{H}_{jk})^{2}] &\simeq& -3y^{2}_{\tau}{\rm Re}[(\widetilde{Y}_{\nu})_{j3}(\widetilde{Y}^{\dag}_{\nu})_{3k}]{\rm Im}[\widetilde{H}_{ij}]\cdot t~~~~~~~~{\rm
 % for}~~\phi_{A}\neq0,~\phi_{B}\neq0,
 %\label{cpcarrying1}
 %\end{eqnarray}

Now, let us consider the case that $\phi_{A}\neq0$ and $\phi_{B}\neq0$ of $\textbf{Y}_{\nu}$ in
Eq. (\ref{FTY1}) at the GUT scale.
In this case, from Eq. (\ref{RG 7}), it is easy to find that  ${\rm Re}[\widetilde{H}_{jk}(0)]=0$
and ${\rm Im}[\widetilde{H}_{jk}(0)]\neq0$, and  thus  RG effects on the off-diagonal elements $\widetilde{H}_{jk}$
may be prominent in the real part as given by
 \begin{eqnarray}
  {\rm Re}[\widetilde{H}_{jk}] \simeq
  -\frac{3}{2}y^{2}_{\tau}{\rm Re}[(\widetilde{Y}_{\nu})_{j3}(\widetilde{Y}^{\dag}_{\nu})_{3k}]\cdot t.
  \label{RGeffect13}
 \end{eqnarray}
With the help of Eqs. (\ref{cpasym0},\ref{RGeffect13}),
the CP-asymmetry can be written as
 \begin{eqnarray}
  \varepsilon_{i} &\simeq&\frac{3y^{2}_{\tau}}{32\pi\widetilde{H}_{ii}}
  \sum_{j}\frac{{\rm Re}[(\widetilde{Y}_{\nu})_{j3}(\widetilde{Y}^{\dag}_{\nu})_{3i}]{\rm Im}[\widetilde{H}_{ji}]}
  {\widetilde{H}_{ii}-\widetilde{H}_{jj}}.
  \label{RGLepto01}
  \end{eqnarray}

In addition to $\varepsilon_{i}$, it is well-known that he baryon
asymmetry depends on the parameters
 \begin{eqnarray}
  K_{i}\equiv\frac{\widetilde{m}_{i}}{\widetilde{m}_{\ast}},~~~~~~~~~ \widetilde{m}_{i}\equiv
  \frac{\widetilde{H}_{ii}}{M_{i}}\upsilon^{2},
  \label{K-factor1}
 \end{eqnarray}
where $m_{\ast}\simeq 10^{-3}eV$ is the so-called equilibrium
neutrino mass and the effective neutrino mass $\widetilde{m}_{i}$
is a measure of the strength of the coupling of $N_{i}$ to the
thermal bath. After reprocessing by sphaleron transitions, the
baryon asymmetry is related to the $(B-L)$ asymmetry by
 $Y_{B}=(12/37)(Y_{B-L})$ \cite{Harvey}. In  flavor independent leptogenesis
we are always in the strong wash-out regime with
$K_{i}\gg1$ and the right-handed neutrinos $N_{i}$`s are nearly in
thermal equilibrium.
Then, the generated $B-L$ asymmetry  in the strong wash-out
regime is given \cite{Abada} as
 \begin{eqnarray}
  Y_{B-L}\simeq
  \sum_{i}0.3\frac{\varepsilon_{i}}{g_{\ast}}\Big(\frac{0.55\times10^{-3} eV}{\widetilde{m}_{i}}\Big)^{1.16},
 \label{convLepto}
 \end{eqnarray}
 where $g_{\ast}$ is the effective number of degrees of freedom.
Therefore, the resulting baryon-to-photon ratio becomes
$\eta_{B}=7.0394\cdot Y_{B}$, where
 \begin{eqnarray}
Y_{B}\simeq
\frac{12}{37}\sum_{i}0.3\frac{\varepsilon_{i}}{g_{\ast}}\Big(\frac{0.55\times10^{-3}
eV}{\widetilde{m}_{i}}\Big)^{1.16}.
 \label{convLepto1}
 \end{eqnarray}
Here the value 7.0394 comes out from the present ratio of entropy density to photon density \cite{Buchmuller:2002rq}.

%%%%%%%%%%%%%%%%%%%%%%%%%%%%%%%%%%%%%%%%%%%%%%%%%%%%%%%%%%%%%%%%%%%%%%%%%%%%%%%%%%%%%%%%%%%%%%%%%%%%%%%%%%%%%%%%%%%%%%%%%%%%%%%%%%%%%%%%%%%%%%%%%%%%%%%%%%%%%%%%%%%%%%%%%%%%%%%%%%%%%%%%%
\subsection{Flavor Dependent Leptogenesis}

Considering flavor effects, the CP asymmetry generated through the
interference between tree and one-loop diagrams of heavy singlet
Majorana neutrino $N_{i}$ decay is given
for each lepton flavor $\alpha(=e,\mu,\tau)$ by \cite{Flavor, SKK} :
 \begin{eqnarray}
  \varepsilon^{\alpha}_{i} &=& \frac{\Gamma(N_{i}\rightarrow l_{\alpha}\varphi)
  -\Gamma(N_{i}\rightarrow \overline{l}_{\alpha}\varphi^{\dag})}{\sum_{\alpha}[\Gamma(N_{i}\rightarrow l_{\alpha}\varphi)
  +\Gamma(N_{i}\rightarrow\overline{l}_{\alpha}\varphi^{\dag})]}\nonumber\\
  &=& \frac{1}{8\pi(Y_{\nu}Y^{\dag}_{\nu})_{ii}}\sum_{j}{\rm
  Im}\Big\{(Y_{\nu}Y^{\dag}_{\nu})_{ij}
  (Y_{\nu})_{i\alpha}(Y_{\nu})^{\ast}_{j\alpha}\Big\}g\Big(\frac{M^{2}_{j}}{M^{2}_{i}}\Big),
 \label{cpasym1}
 \end{eqnarray}
where $j$ runs over 1, 2 and 3 but $i\neq j$ and the function
$g(M^{2}_{j}/M^{2}_{i})$ is given by Eq. (\ref{decayfunction}). We
note that the total CP asymmetries $\varepsilon_{i}$ in Eq.
(\ref{cpasym0}) are obtained by summing over the lepton flavors
$\alpha$. {}From Eq. (\ref{cpasym1}), we see that leptogenesis
reflecting flavor effects depends not only on
$Y_{\nu}Y^{\dag}_{\nu}$ but also on the individual $Y_{\nu}$, which
makes it different from the conventional leptogenesis. The CP
asymmetry $\varepsilon^{\alpha}_{i}$ is resonantly enhanced when the
decay width of the $j$th-generation right-handed neutrino
$\Gamma_{j}\simeq(M^{2}_{i}-M^{2}_{j})/M_{i}$. Once the initial
values of $\varepsilon^{\alpha}_{i}$ are fixed, the final result of
$\eta_{B}$ or $Y_{B}$ will be governed by a set of flavor-dependent
Boltzmann equations including the (inverse) decay and scattering
processes as well as the nonperturbative sphaleron interaction
\cite{Flavor,Barbieri,PU2}.

In the case of $\phi_{A}=\phi_{B}=0$  at the GUT scale, the CP-asymmetry of a single
flavor $\alpha$ including RG effects from high-energy scale to seesaw scale
is approximately written as
 \begin{eqnarray}
  \varepsilon^{\alpha}_{i} &\simeq& \frac{3y^{2}_{\tau}}{32\pi\widetilde{H}_{ii}}
  \sum_{j}\frac{{\rm Im}[(\widetilde{Y}_{\nu})_{i3}(\widetilde{Y}^{\ast}_{\nu})_{j3}]{\rm Re}[(\widetilde{Y}_{\nu})_{i\alpha}(\widetilde{Y}^{\ast}_{\nu})_{j\alpha}]}
  {\widetilde{H}_{jj}-\widetilde{H}_{ii}}\nonumber\\
  &&+\frac{3y^{2}_{\tau}}{64\pi\widetilde{H}_{ii}}
  \sum_{j}\frac{{\rm Re}[(\widetilde{Y}_{\nu})_{i3}(\widetilde{Y}^{\ast}_{\nu})_{j3}]{\rm Im}[(\widetilde{Y}_{\nu})_{i\alpha}(\widetilde{Y}^{\ast}_{\nu})_{j\alpha}]}
  {\widetilde{H}_{jj}-\widetilde{H}_{ii}}.
 \label{RGLepto}
 \end{eqnarray}
Here, we note that $(Y_{\nu})_{i\alpha}(Y^{\dag}_{\nu})_{\alpha i}$ contains
the CP-phases $\varphi_{A}$ and $\varphi_{B}$  which may enhance
the CP asymmetry.

 On the other hand, in the case of $(\phi_{A}\neq0,
 \phi_{B}\neq0)$  at the GUT scale, the imaginary part in Eq. (\ref{cpasym1}) including RG effects  becomes
 \begin{eqnarray}
  &&{\rm Im}\Big\{\widetilde{H}_{jk}(Y_{\nu})_{j\alpha}(Y_{\nu})^{\ast}_{k\alpha}\Big\}\nonumber\\
   &\simeq& {\rm Im}[\widetilde{H}_{jk}]{\rm Re}[(\widetilde{Y}_{\nu})_{j\alpha}(\widetilde{Y}^{\ast}_{\nu})_{k\alpha}]
   -\frac{3y^{2}_{\tau}}{2}{\rm Re}[(\widetilde{Y}_{\nu})_{j3}(\widetilde{Y}^{\ast}_{\nu})_{k3}]{\rm Im}
   [(\widetilde{Y}_{\nu})_{j\alpha}(\widetilde{Y}^{\ast}_{\nu})_{k\alpha}]\cdot t~~~ (j\neq k),
 \label{cpcarrying4}
 \end{eqnarray}
where the first term in the second line dominates over the second
one. We see from Eq. (\ref{cpcarrying4}) that CP asymmetry can be
generated without the CP phases $\varphi_{A,B}$ in this case.
Neglecting the second term in Eq. (\ref{cpcarrying4}), the
CP-asymmetry of a single flavor $\alpha$ is approximately written
as
 \begin{eqnarray}
  \varepsilon^{\alpha}_{i}
  \simeq \frac{\pi}{2\widetilde{H}_{ii}\cdot{\rm ln}(M_{i}/\Lambda)}\sum_{j}\frac{{\rm Re}
  [(\widetilde{Y}_{\nu})_{i\alpha}(\widetilde{Y}^{\ast}_{\nu})_{j\alpha}]
  {\rm Im}[\widetilde{H}_{ij}]}{\widetilde{H}_{ii}-\widetilde{H}_{jj}}.
 \label{RGLepto1}
 \end{eqnarray}

In order to estimate the washout effects, one
may introduce the parameter $K^{\alpha}_{i}$ which is the washout
factor due to the inverse decay of the Majorana neutrino $N_{i}$
into the lepton flavor $\alpha(=e,\mu,\tau)$ \cite{Abada}
 \begin{eqnarray}
  K^{\alpha}_{i} =\frac{\Gamma(N_{i}\rightarrow l_{\alpha}\varphi)
  +\Gamma(N_{i}\rightarrow \overline{l}_{\alpha}\varphi^{\dag})}{\sum_{\alpha}[\Gamma(N_{i}\rightarrow l_{\alpha}\varphi)
  +\Gamma(N_{i}\rightarrow\overline{l}_{\alpha}\varphi^{\dag})]}K_{i}
  = \frac{(Y_{\nu})_{i\alpha}(Y^{\dag}_{\nu})_{\alpha i}}{(Y_{\nu}Y_{\nu}^{\dag})_{ii}}K_{i},
  \label{K-factor2}
 \end{eqnarray}
where
 \begin{eqnarray}
  K_{i}=\sum_{\alpha=e,\mu,\tau}K^{\alpha}_{i}=\frac{\Gamma_{i}}{H(T=M_{i})},~~~K^{\alpha}=\sum_{i=1}^{3}K^{\alpha}_{i},
 \end{eqnarray}
with $\Gamma_{i}=\sum_{\alpha}\Gamma^{\alpha}_{i}$ denoting the
total decay width of $N_{i}$ at tree level where
$\Gamma^{\alpha}_{i}$ is the partial decay rate of the process
$N_{i}\rightarrow l_{\alpha}+\varphi^{\dag}$. The washing out of a
given flavor $l_{\alpha}$ is operated by the $\Delta L=1$
scattering involving all three right-handed neutrinos, which is
parameterized by
\begin{eqnarray}
  \widetilde{m}^{\alpha}_{i}=(Y^{\dag}_{\nu})_{\alpha
  i}(Y_{\nu})_{i\alpha}\frac{\upsilon^{2}}{M_{i}},~~~~~~~~~~~~~~~~
  \frac{\widetilde{m}^{\alpha}_{i}}{m_{\ast}}=\frac{\Gamma(N_{i}\rightarrow \varphi l_{\alpha})}{H(M_{i})},
\end{eqnarray}
where $\widetilde{m}^{\alpha}_{i}$ parameterizes the decay rate of
$N_{i}$ to the leptons of flavor $l_{\alpha}$ and the trace
$\sum_{\alpha}\widetilde{m}^{\alpha}_{i}$ coincides with the
$\widetilde{m}_{i}$ parameter defined in the previous section. The
each lepton asymmetries are washed out differently by the
corresponding washout parameter which is given by Eq.
(\ref{K-factor2}), and appear with different weights in the final
formula for the baryon asymmetry \cite{Abada}, as will be shown
later (see Eqs. (\ref{baryon2}-\ref{baryon3})). Indeed the lepton
asymmetry for each flavor $l_{\alpha}$  generated through $N_{i}$
decay is given by
 \begin{eqnarray}
  Y^{\alpha}_{i} &\simeq&
  0.3\frac{\varepsilon^{\alpha}_{i}}{g_{\ast}}\Big(\frac{0.55\times10^{-3} eV}{\widetilde{m}^{\alpha}_{i}}\Big)^{1.16}
 \label{strong}
 \end{eqnarray}
 in the strong wash-out regime ( $K^{\alpha}_{i}\gg1$), and
  \begin{eqnarray}
  Y^{\alpha}_{i} &\simeq&
  1.5\frac{\varepsilon^{\alpha}_{i}}{g_{\ast}}\Big(\frac{\widetilde{m}_{i}}{3.3\times10^{-3} eV}\Big)\Big(\frac{\widetilde{m}^{\alpha}_{i}}{3.3\times10^{-3} eV}\Big)
  \label{weak}
 \end{eqnarray}
 in the weak wash-out regime ( $K^{\alpha}_{i}\ll1$).

For temperatures $10^{9}~{\rm GeV}\lesssim T\sim M_{i}\lesssim
10^{12}~{\rm GeV}$, the interactions mediated by the $\tau$ Yukawa
coupling are in equilibrium, whereas those by the other Yukawa
couplings are out of equilibrium. Then, the lepton asymmetries for
the electron and muon flavors can be treated as a linear
combination: $Y^{2}_{i}\equiv Y^{e}_{i}+Y^{\mu}_{i}$. Finally, the
baryon asymmetry is given by
    \cite{Abada}
\begin{eqnarray}
     Y_{\emph{B}}\simeq
     \frac{12}{37}\sum_{N_{i}}\Big[Y^{2}_{i}\Big(\varepsilon^{2}_{i},\frac{417}{589}\widetilde{m}_{2}\Big)
     +Y^{\tau}_{i}\Big(\varepsilon^{\tau}_{i},\frac{390}{589}\widetilde{m}^{\tau}_{i}\Big)\Big],
    \label{baryon2}
    \end{eqnarray}
where $\varepsilon^{2}_{i}=\varepsilon^{e}_{i}+\varepsilon^{\mu}_{i}$, and the corresponding wash-out parameter is
    $K^{2}_{i}=K^{e}_{i}+K^{\mu}_{i}$.

Below temperatures $T\sim M_{i}\lesssim10^{9}$ GeV, muon and tau
charged lepton Yukawa interactions are much faster than the Hubble
expansion parameter rendering the $\mu$ and $\tau$ Yukawa couplings in
equilibrium.
 Then, in this case the final baryon asymmetry is given \cite{Abada} as
\begin{eqnarray}
     Y_{\emph{B}}\simeq
     \frac{12}{37}\sum_{N_{i}}\Big[Y^{e}_{i}\Big(\varepsilon^{e}_{i},\frac{151}{179}\widetilde{m}^{e}_{i}\Big)
     +Y^{\mu}_{i}\Big(\varepsilon^{\mu}_{i},\frac{344}{537}\widetilde{m}^{\mu}_{i}\Big)
     +Y^{\tau}_{i}\Big(\varepsilon^{\tau}_{i},\frac{344}{537}\widetilde{m}^{\tau}_{i}\Big)\Big].
    \label{baryon3}
    \end{eqnarray}
Notice that the CP-asymmetries of a single flavor given in  Eqs. (\ref{baryon2},\ref{baryon3}) are
weighted separately due to the different values of $\widetilde{m}^{\alpha}_{i}$.

In the strong washout regime, which corresponds to our case, given
the initial thermal abundance of $N_{i}$ and the condition
$K^{\alpha}_{i}\gtrsim1$, the baryon asymmetry including lepton
flavor effects is given \cite{PU2} as
  \begin{eqnarray}
   \eta_{B}\simeq-0.96\times10^{-2}\sum_{i}\sum_{\alpha}\varepsilon^{\alpha}_{i}\frac{K^{\alpha}_{i}}{K_{i}K^{\alpha}}~.
   \label{etaB}
  \end{eqnarray}
The ratio of $\eta_B$, generated through flavor independent
leptogenesis,  to $\eta^{\rm flavor}_B$, generated through flavor
dependent leptogenesis,
%where the dominant contribution to these estimates comes from $N_{3}$, and taking into account of Eq. (\ref{etaB}), we demonstrate that the
%ratio of the two estimates
in $10^{9}~{\rm GeV}\lesssim  M \lesssim 10^{12}~{\rm GeV}$  region yields
  \begin{eqnarray}
   \frac{\eta_{B}}{\eta^{\rm
   flavor}_{B}}\sim\frac{\varepsilon_{3}}{\varepsilon^{\tau}_{3}}\frac{K^{\tau}}{K^{\tau}_{3}}\approx\frac{y^{2}_{\tau}}{8\pi^{2}}{\rm
   ln}\big(\frac{M}{\Lambda}\big)\frac{K^{\tau}}{K^{\tau}_{3}},
  \end{eqnarray}
  where the orders of magnitude of $K^{\tau}$ and $K^{\tau}_{3}$ are $\sim
  O(100)$ and  $\sim O(1)$, respectively. Thus, without taking lepton flavor effects into account, in
this region  the prediction of $\eta_B$ is
suppressed by $4\sim 5$ orders of magnitude compared with $\eta^{\rm flavor}_B$.

Below the temperature
$M\sim 10^{9}$ GeV, all cases of parameter spaces can contribute to
leptogenesis with different washout-factors. As indicated in Eqs.
(\ref{RGLepto00},\ref{RGLepto1}), the CP-asymmetries
$\varepsilon_{i}$ and $\varepsilon^{\alpha}_{i}$ are weakly
dependent on the heavy Majorana neutrino scale $M$. Without taking
account of wash-out factors, since there is no CP-violation phases
at the degeneracy scale,
in this case we can obtain approximately $\varepsilon_{i}\propto y^{4}_{\tau} t$
and $\varepsilon^{\alpha}_{i}\propto y^{2}_{\tau}$, see Eqs. (\ref{RGLepto00},\ref{RGLepto}),
and the CP-asymmetry
$\varepsilon^{\alpha}_{i}$ gets enhanced
by $\varepsilon^{\alpha}_{i}/\varepsilon_{i}\sim 1/y^{2}_{\tau}t$ due to flavor effects.

%%%%%%%%%%%%%%%%%%%%%%%%%%%%%%%%%%%%%%%%%%%%%%%%%%%%%%%%%%%%%%%%%%%%%%%%%%%%%%%%%%%%%%%%%%%%%%%%%%%%%
\section{Numerical analysis}
%%%%%%%%%%%%%%%
%    Fig 5    %
%%%%%%%%%%%%%%%
\begin{figure}[b]
%\vspace*{-5.0cm}
\hspace*{-2cm}
\begin{minipage}[t]{6.0cm}
\epsfig{figure=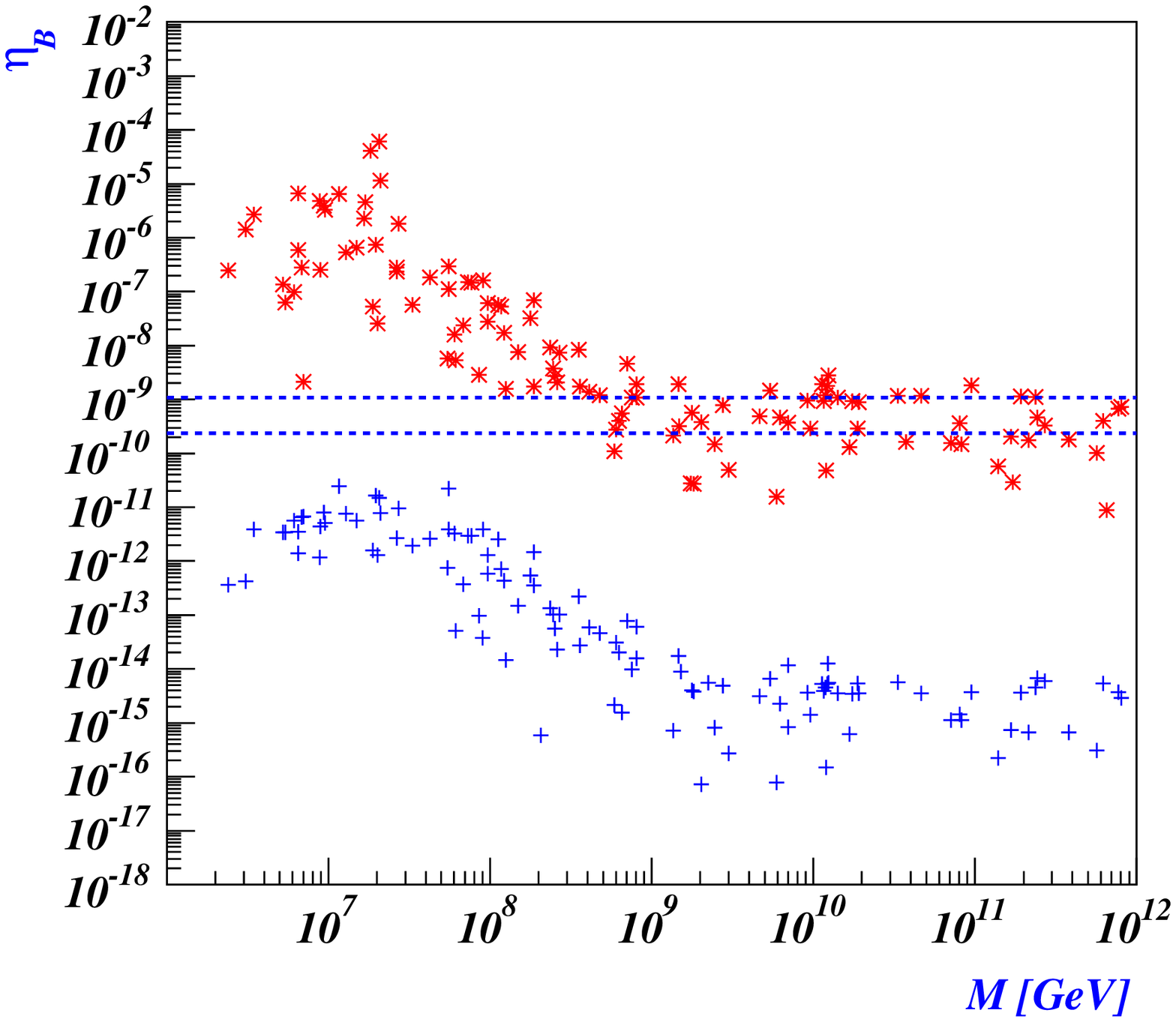,width=6.5cm,angle=0}
\end{minipage}
\hspace*{1.0cm}
\begin{minipage}[t]{6.0cm}
\epsfig{figure=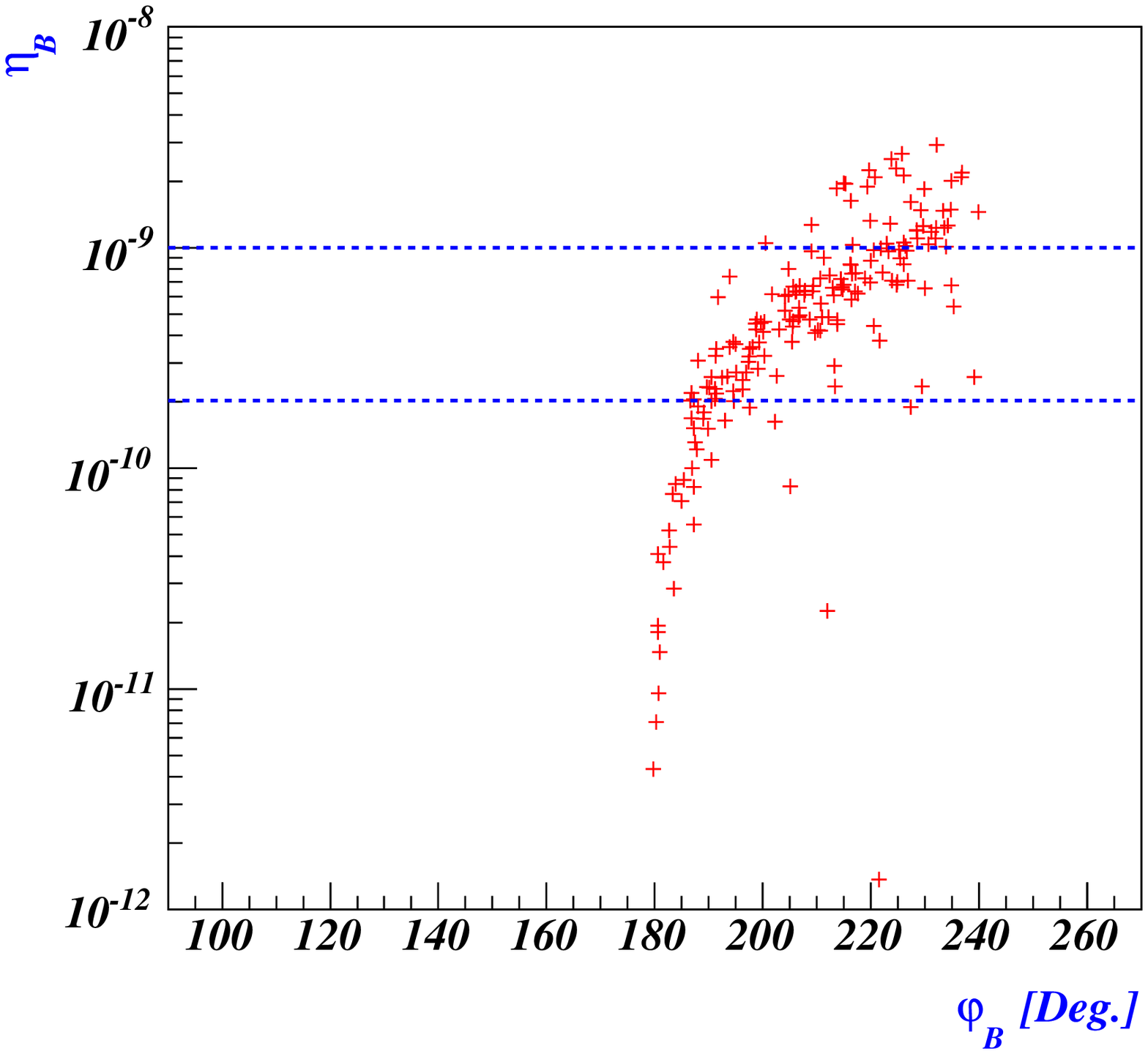,width=6.5cm,angle=0}
\end{minipage}
%\vspace*{-5.5cm}
\caption{ \label{Fig5} Left-figure shows the predictions of
$\eta_B$ for $10^{6}\lesssim M[{\rm GeV}]\lesssim 10^{12}$. The
asters correspond to flavored leptogenesis and the crosses
correspond to flavor independent leptogenesis. Right-figure shows
$\eta^{\rm flavor}_{B}$ as a function of $\varphi_{B}$ for
$10^{9}\lesssim M[{\rm GeV}]\lesssim 10^{12}$. The horizontal
dotted lines in both figures correspond to the current measurement
from WMAP \cite{cmb}.}
\end{figure}

%For a successful leptogenesis at the seesaw scale in addition to
%$\delta_{N}\neq0$ (the degree of degeneracy of heavy Majorana
%neutrino masses), the Yukawa Dirac coupling matrices have to
%contain complex phases.
Confronting neutrino masses and mixing in the context of our
scheme with low energy neutrino experimental data given in Eq.
(\ref{exp bound}), we determine the allowed regions of the model
parameters for which we estimate the lepton asymmetry. For the
case of $\phi_{A}=\phi_{B}=0$ at the GUT scale, in left figure
of Fig. 5, we plot the predictions of baryon asymmetry $\eta_B$
for $10^{6}\lesssim M[{\rm GeV}]\lesssim10^{12} $. The horizontal
dotted lines correspond to the bounds on $\eta_B$ measured from
current astrophysical observations, ($2\times
10^{-10}<\eta_B<10\times 10^{-10}$). The asters correspond to
 flavored leptogenesis, whereas the crosses correspond to
 flavor independent leptogenesis. We see from left figure of
Fig.~\ref{Fig5} that successful leptogenesis in the FTY model is
possible only when lepton flavor effects are included, and the
required values of $\eta_B$ can be achievable for the temperature
ranges of $M\gtrsim10^{9}$ GeV. As explained before, for
$10^{9}\lesssim M[{\rm GeV}]\lesssim10^{12}$, only the
interactions mediated by the $\tau$ Yukawa coupling are in
equilibrium and thus only the $\tau$-flavor is treated separately
in the Boltzmann equations while the $e$ and $\mu$ flavors are
indistinguishable. Left-figure of Fig. 5 shows that FTY structure
reaches maximal $\eta_{B}$ near $10^{7}$ GeV (seesaw scale)
running down from GUT scale, corresponding to $M_{1}\lesssim
M_{2}\lesssim M_{3}$, which is related with the stable angle
$\theta_{i}$ in Eq. (\ref{RG 0}) (see also \cite{RG1}).

%%%%%%%%%%%%%%%
%    Fig6     %
%%%%%%%%%%%%%%%
\begin{figure}[t]
%\vspace*{-5.0cm}
\hspace*{-2cm}
\begin{minipage}[t]{6.0cm}
\epsfig{figure=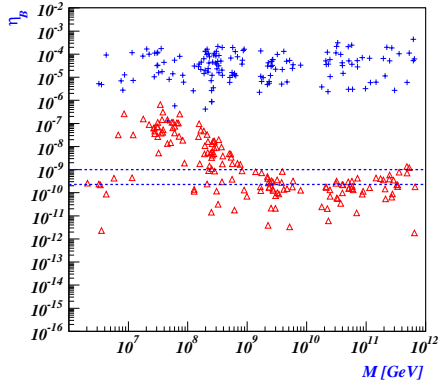,width=6.5cm,angle=0}
\end{minipage}
%\vspace*{-5.5cm}
\caption{\label{Fig6} The predictions of the BAU $\eta_{B}$ for
$10^{6}\lesssim M[{\rm GeV}]\lesssim 10^{12}$. The horizontal
dotted lines correspond to the current observation from WMAP
\cite{cmb}. The crosses correspond to flavored leptogenesis and
the triangles correspond to flavor independent leptogenesis.}
\end{figure}
For $10^{9}\lesssim M[{\rm GeV}]\lesssim10^{12}$, right figure of
Fig.~\ref{Fig5} represents how the predictions of $\eta_{B}$ in
flavored leptogenesis depend on the initial value of the phase
$\varphi_{B}$ imposed at GUT scale. In the same region of $M$, we
find that $\eta^{\tau}_B$ dominates over
$\eta^{2}_{B}=\eta^{e}_{B}+\eta^{\mu}_{B}$, and thus the
successful leptogenesis in the FTY model is approximately equal to
tau-resonant leptogenesis \cite{PU2}.

In the case of $\phi_{A}\neq0$ and $\phi_{B}\neq0$ at the GUT
scale,  Fig.~\ref{Fig6} presents the predictions of $\eta_B$
generated through flavor independent leptogenesis (the triangles)
and those of $\eta^{\rm flavor}_B$ through flavor dependent
leptogenesis (the crosses) for $10^{6}\lesssim M[{\rm
GeV}]\lesssim10^{12}$. Note that we vary the values of
$\phi_{A,B}$ as well as $\varphi_{A,B}$ from 0 to $2\pi$ without
fixing certain values. The horizontal dotted lines correspond to
the current bounds on $\eta_B$. We see from Fig.~\ref{Fig6} that
flavor independent leptogenesis leads to the right amount of
baryon asymmetry required from the current observational result,
whereas the predictions for $\eta^{\rm flavor}_B$ are too large
for flavor dependent leptogenesis to be a desirable candidate for
baryogenesis. The reason  why $\eta_B^{\rm flavor}$ get enhanced
compared with $\eta_B$ generated through flavor independent
leptogenesis is that the first contribution in Eq.
(\ref{cpcarrying4}) dominates over the second one, so that
$\varepsilon^{\alpha}_{i}/\varepsilon_{i}\sim 1/y^{2}_{\tau}t$
which is much less than one.

%%%%%%%%%%%%%%%%%%%%%%%%%%%%%%%%%%%%%%%%%%%%%%%%%%%%%%%%%%%%%%%%%%%%%%%%%%%%%%%%%%%%%%%%%%%%%%%%%%%%%
\section{summary}

As a summary, we have considered FTY model \cite{FTY} realized at
the GUT scale. By considering RG evolution from GUT scale to low
energy scale, we have confronted light neutrino masses and mixing
with low energy experimental data, and  found the allowed
parameter space. We have investigated how BAU
can be achieved via leptogenesis in FTY model. In
particular, we considered two scenarios, one is to include lepton
flavor effects and the other is to ignore them. In FTY model we
consider, there are two types of CP phases, $\phi_{A,B}$ appeared
in $Y_{\nu}$ and $\varphi_{A,B}$ in $Y_{l}$. Besides those CP
phases, we need to splitting of the heavy Majorana neutrino
spectrum in order to generate lepton asymmetry in FTY model. We
have shown that the desirable splitting of the heavy Majorana
neutrino  spectrum could be radiatively induced at the seesaw
scale by using the RG evolution from GUT to seesaw scale. In the
case of $\phi_{A}=0, \phi_{B}=0$  at GUT scale, we have found that
the predictions of $\eta_B$ through flavor independent
leptogenesis are not enough to achieve successful baryogenesis,
whereas it can be achieved by flavor dependent leptogenesis for
$10^{9}\lesssim M[\rm GeV]\lesssim10^{12}$.

In the case of the phases $\phi_{A}\neq0, \phi_{B}\neq0$ at the
GUT scale, contrary to the previous case, the successful
leptogenesis can be achieved by ignoring the lepton flavor effects
because flavor effects greatly enhance the lepton asymmetry so
that they are not desirable to achieve baryon asymmetry of our
universe.

We note that in both cases of our work, leptogenesis can be viable for
  \begin{eqnarray}
10^{9}\lesssim M[\rm GeV]\lesssim10^{12}.
  \end{eqnarray}
In particular, in the FTY model, flavor dependent leptogenesis can
be worked when $Y_{\nu}$ does not contain  CP-phases, but $Y_l$
contains CP-phases.

%%%%%%%%%%%%%%%%%%%%%%%%%%%%%%%%%%%%%%%%%%%%%%%%%%%%%%%%%%%%%%%%%%%%%%%%%%%%%%%%%%%%%%%%%%%%%%%%%%%%%%%%%%
\acknowledgments{ \noindent YHA was supported  by the Korea Research Foundation
Grant funded by the Korean Government (MOEHRD) No.
KRF-2005-070-C00030. CSK was supported in part by  CHEP-SRC
Program and in part by the Korea Research Foundation Grant funded
by the Korean Government (MOEHRD) No. KRF-2005-070-C00030.
JL was supported in part by Brain Korea 21 Program
and in part by Grant No. F01-2004-000-10292-0 of KOSEF-NSFC
International Collaborative Research Grant. SKK was supported
 by KRF Grant funded by the Korean Government (MOEHRD) No. KRF-2006-003-C00069.}

%%%%%%%%%%%%%%%%%%%%%%%%%%%%%%%%%%%%%%%%%%%%%%%%%%%%%%%%%%%%%%%%%%%%%%%%%%%%%%%%%%%%%%%%%%%%%%%%%%%%%%%%%%%

\end{document}